\documentclass[reprint,
preprintnumbers,
nofootinbib,
amsmath,amssymb,
aps,
prd,
superscriptaddress,
]{revtex4-2}

\usepackage{graphicx}
\usepackage{dcolumn}
\usepackage{siunitx}
\usepackage{bm}
\usepackage[hidelinks]{hyperref}
\usepackage[all]{hypcap}
\usepackage{xcolor}
\usepackage{enumitem}
\usepackage{media9}
\usepackage{animate}
\usepackage{multirow}
\usepackage{orcidlink}

\hypersetup{
	colorlinks,
	linkcolor={red!50!black},
	citecolor={blue!50!black},
	urlcolor={blue!80!black}
}

\newcommand{\order}[1]{\mathcal{O}(#1)}

\DeclareMathOperator{\Tr}{Tr}
\renewcommand{\Re}{\operatorname{Re}}

\makeatletter
\def\maketitle{
	\@author@finish
	\title@column\titleblock@produce
	\suppressfloats[t]}
\makeatother

\begin{document}

\preprint{ADP-25-5/T1267}

\title{Structure of center-vortex matter in SU(4) Yang-Mills theory}

\author{Jackson A. Mickley \orcidlink{0000-0001-5294-2823}}
\affiliation{Centre for the Subatomic Structure of Matter, Department of Physics, The University of Adelaide, South Australia 5005, Australia}

\author{Derek B. Leinweber \orcidlink{0000-0002-4745-6027}}
\affiliation{Centre for the Subatomic Structure of Matter, Department of Physics, The University of Adelaide, South Australia 5005, Australia}

\author{Luis E. Oxman \orcidlink{0000-0002-4670-5597}}
\affiliation{Instituto de F{\'i}sica, Universidade Federal Fluminense, 24210-346 Niter{\'o}i, RJ, Brasil}

\begin{abstract}
The structure of center vortices is studied in $\mathrm{SU}(4)$ Yang-Mills theory for the first time to illuminate the interplay between elementary (center charge $\pm 1$) and doubly charged vortices. Unlike in $\mathrm{SU}(3)$, where charge $+2$ vortices are simply elementary vortices with reversed orientations in spacetime, these possibilities are physically distinct in $\mathrm{SU}(4)$. Visualizations of the vortex structure in three-dimensional slices reveal the various ways in which doubly charged objects manifest, as the convergence and matching of elementary vortices or as isolated doubly charged loops. An algorithm is described to classify every doubly charged chain as one of these three types. A collection of vortex statistics is considered to quantify the vortex structure. Many of these pertain to the novel doubly charged objects, including their relative proportions and chain lengths, which are analyzed to highlight the differences between each chain type. Three different lattice spacings are employed to investigate the approach to the continuum limit. Vortex matching chains are found to be shorter on average but also more prevalent than vortex convergences, ascribed to their interpretation as extended center monopoles. In addition, the lengths of both vortex convergences and vortex matchings are observed to follow an exponential distribution, allowing the introduction of a constant probability for a doubly charged chain to split into two elementary vortices as it propagates. Combined, these findings provide a characterization of the vortices that comprise center-vortex structures in $\mathrm{SU}(4)$ Yang-Mills theory.
\end{abstract}

\maketitle

\section{Introduction} \label{sec:intro}
Over the past five decades, it has become increasingly clear that a quantum vacuum dominated by topological objects such as center vortices is essential to understanding the confinement mechanism in 4D $\mathrm{SU}(N)$ Yang-Mills (YM) theory \cite{Mandelstam:1974pi, tHooft:1977nqb, tHooft:1979rtg, Mack:1978rq, Nielsen:1979xu, DelDebbio:1996lih, DelDebbio:1997ep, Langfeld:1997jx, DelDebbio:1998luz, Faber:1997rp, Faber:1998qn, Kovacs:1998xm, Langfeld:1998cz, Bertle:1999tw, Engelhardt:1999fd, Engelhardt:1999wr, Faber:1999sq, deForcrand:1999our, Kovacs:2000sy, Langfeld:2001cz, Langfeld:2003ev, Greensite:2003bk, Engelhardt:2003wm, Bowman:2008qd, Bowman:2010zr, OMalley:2011aa, Greensite:2016pfc, Biddle:2022zgw, Biddle:2022acd, Mickley:2024zyg, Mickley:2024vkm}. These physical objects are part of a complex landscape that may have a large spectrum of possibilities. Further progress requires mapping this landscape across various scenarios, connecting it with potential effective descriptions, and establishing links with different observables. In this way, a coherent unified picture of confinement can start to take shape.

Early seminal works laid the foundations for understanding the key role of center vortices in confinement \cite{Mandelstam:1974pi, tHooft:1977nqb, tHooft:1979rtg, Mack:1978rq, Nielsen:1979xu}. One of the methods that made their detection possible is based on a two-step process. Initially, the Monte Carlo configurations are transformed to physically equivalent ones in the direct maximal center gauge (DMC), where the link variables are as close as possible to center elements. Next, center-projected variables are used to detect plaquettes with nontrivial center charge, which are pierced by the center-vortex guiding centers. In $\mathrm{SU}(2)$ and $\mathrm{SU}(3)$, this procedure showed that center-vortex world surfaces percolate the full four-dimensional spacetime at low temperatures \cite{DelDebbio:1996lih, DelDebbio:1998luz, Engelhardt:1999fd}.

When compared with $\mathrm{SU}(2)$, a new possibility emerges in $\mathrm{SU}(3)$, with center-vortex world surfaces able to be matched in groups of three. This property was studied and characterized with or without dynamical quarks by following them over spacetime slices \cite{Langfeld:2003ev, Spengler:2018dxt, Biddle:2019gke, Biddle:2023lod, Mickley:2024zyg, Mickley:2024vkm}. In pure $\mathrm{SU}(3)$ YM at zero temperature, a network of connected percolating center-vortex lines forming a primary cluster and secondary loops that grow and ``merge'' with the primary cluster were observed \cite{Biddle:2019gke, Mickley:2024zyg}. The matchings are due to center-charge conservation modulo 3, which allows for three charge $+1$ vortex lines to end (or start) at a common point. This can also be thought of as a pair of charge $+1$ vortices that converge and fuse into a charge $+2$ vortex, as this gives an elementary percolating vortex:\ $+2 = -1 \pmod 3$.

In the context of $\mathrm{SU}(4)$ gauge theory, the physical inequivalence between elementary and doubly charged center vortices adds another layer of complexity. Four charge $+1$ center-vortex lines could match at a common point (center monopole) or through an extended object, i.e.\ an intermediate doubly charged vortex line. Also, a pair of charge $+1$ vortices could converge to form a doubly charged line and then split. At the commencement of this study, there was no knowledge from \textit{ab initio} Yang-Mills simulations about whether these intermediate components are favored or not, nor about the behavior of elementary and doubly charged secondary loops. Indications to the relevance of these intermediate doubly charged objects has previously been indirectly inferred from phenomenological studies \cite{Engelhardt:2005qu}.

Center vortices have also been studied and discussed with methods based on the maximal Abelian gauge that can additionally detect the presence of monopoles on them, with nontrivial Cartan flux \cite{DelDebbio:1997ke, Ambjorn:1999ym, Alexandrou:1999vx, deForcrand:2000pg}. The chains formed when monopoles are present, also known as nonoriented center vortices, contribute to the topological charge \cite{Reinhardt:2001kf}. Furthermore, when mixed with oriented center vortices, these chains can explain the formation of the confining flux tube and its properties at asymptotically large distances between quarks \cite{Oxman:2018dzp, Junior:2022bol, Junior:2024urr}.

The main objective of this work is to determine the general structure of center-vortex networks in $\mathrm{SU}(4)$ lattice YM theory and reveal details about the different components observed. As these studies will be based on the DMC gauge, the network will be formed by all possible oriented and nonoriented center vortices, with no information about the Cartan flux orientations that distinguish them. The main focus will be on elucidating how vortex convergences (VC), vortex matchings (VM) and doubly charged loops (L2) are realized. To achieve this, high-resolution simulations on fine lattices, accompanied by a detailed statistical analysis, will be implemented. This characterization will provide novel features of the $\mathrm{SU}(4)$ network as compared to $\mathrm{SU}(3)$, which will help to confirm or refine effective proposals for the confinement mechanism in $\mathrm{SU}(N)$ pure YM theory. 

This paper is structured as follows. We start by summarizing the center-vortex model and the procedure for identifying vortices in Sec.~\ref{sec:centervortices}. Visualizations of $\mathrm{SU}(4)$ vortex structures and the algorithm employed to identify their various components are presented in Sec.~\ref{sec:visualizations}. Section \ref{sec:analysis} contains our quantitative analysis of vortex statistics, including extrapolations to the continuum limit. A comparison is drawn to the random vortex world-surface model for $\mathrm{SU}(4)$ \cite{Engelhardt:2005qu}. Finally, we conclude our primary findings in Sec.~\ref{sec:conclusion}. Supplemental material providing interactive models and embedded animations of $\mathrm{SU}(4)$ center-vortex structures is located at the end of this document. Instructions on interacting with this content is given therein, and the figures are referenced in the main text as Fig.~S-x.

\section{Center vortices} \label{sec:centervortices}
Center vortices~\cite{Mandelstam:1974pi, tHooft:1977nqb, tHooft:1979rtg, Mack:1978rq, Nielsen:1979xu} are regions of the gauge field that carry magnetic flux quantized according to the center of $\mathrm{SU}(N)$,
\begin{equation}
	\mathbb{Z}_N = \left\{ \exp\left(\frac{2\pi i}{N}\, n \right) \mathbb{I} \;\middle|\; n = 0,\, 1,\, \hdots,\, N-1 \right\} \,.
\end{equation}
Physical vortices in the ground-state fields of $\mathrm{SU}(N)$ Yang-Mills theory have a finite thickness. In contrast, on the lattice ``thin" center vortices are extracted through a well-known gauge-fixing procedure that seeks to bring each link variable $U_\mu(x)$ as close as possible to an element of $\mathbb{Z}_N$. These thin vortices form closed surfaces in four-dimensional Euclidean spacetime, and thus one-dimensional structures in a three-dimensional slice of the four-dimensional spacetime.

The gauge fixing is typically performed by finding the gauge transformation $\Omega(x)$ to maximize the below functional, which corresponds to implementing the ``direct" maximal center gauge \cite{Montero:1999by},
\begin{equation}
	R = \sum_{x,\,\mu} \,\left| \Tr U_\mu^{\Omega}(x) \right|^2 \,.
\end{equation}
The links are subsequently projected onto the center,
\begin{equation}
	U_\mu^{\Omega}(x) \longrightarrow Z_\mu(x) = \exp\left(\frac{2\pi i}{N} \, n_\mu(x) \right) \mathbb{I} \in \mathbb{Z}_N \,,
\end{equation}
with $n_\mu(x) \in \{0,\, 1,\, \hdots,\, N-1\}$ identified as the center phase nearest to $\arg \Tr U_\mu(x)$ for each link. Finally, the locations of vortices are identified by nontrivial plaquettes in the center-projected field,
\begin{equation} \label{eq:centerprojplaq}
	P_{\mu\nu}(x) = \prod_\square Z_\mu(x) = \exp\left(\frac{2\pi i}{N} \, m_{\mu\nu}(x) \right)\mathbb{I}
\end{equation}
with $m_{\mu\nu}(x) \neq 0$. The value of $m_{\mu\nu}(x)$ is referred to as the \textit{center charge} of the vortex, and we say the plaquette is pierced by a vortex.

Due to a Bianchi identity satisfied by the vortex fields~\cite{Engelhardt:1999wr, Spengler:2018dxt}, the center charge is conserved such that the vortex topology constitutes closed arrays of sheets in four dimensions, or closed networks of lines in three-dimensional slices of the lattice. Although this procedure is gauge dependent, numerical evidence strongly suggests that the projected vortex locations correspond to the physical guiding centers of thick vortices in the original field configurations~\cite{DelDebbio:1998luz, Langfeld:2003ev, Montero:1999by, Faber:1999gu}. This allows one to investigate the significance of center vortices through the vortex-only field $Z_\mu(x)$.

Our simulations are performed using a pseudo-heat-bath \cite{Creutz:1980zw} algorithm with the standard Wilson action \cite{Wilson:1974sk},
\begin{equation}
	S_W = \beta \sum_{x,\,\mu<\nu} \left[1 - \frac{1}{N} \Re\Tr P_{\mu\nu}(x) \right] \,,
\end{equation}
where $\beta = 2N/g^2$ and $g$ is the bare gauge coupling. The Cabibbo-Marinari algorithm \cite{Cabibbo:1982zn} is applied by iterating over the $N(N-1)/2 = 6$ (for $N = 4$) diagonal $\mathrm{SU}(2)$ subgroups twice per local update.

100 configurations are generated at three different values of $\beta$ to investigate the lattice-spacing dependence of our calculations, with the lattice dimensions varied to keep the physical volume approximately constant. The scale is set using string tension data and the interpolating formula from Refs.~\cite{Lucini:2004my, Lucini:2005vg}. Simulation details are provided in Table~\ref{tab:simdetails}.
\begin{table}
	\centering
	\caption{\label{tab:simdetails} The simulation details for our ensembles, including the $\beta$ value, lattice spacing $a$ in terms of the string tension and in physical units, and the number of spatial $N_s$ and temporal $N_t$ lattice sites. The physical lattice spacing is obtained using for the string tension $\sqrt{\sigma} \simeq 465\,\text{MeV}$ \cite{Bali:1997am, Edwards:1997xf, Allton:2008ty, Athenodorou:2020ani}}
	\begin{ruledtabular}
		\begin{tabular}{ccccc}
			$\beta$ & $a$ ($\sigma^{-1/2}$) & $a$ (fm) & $N_s$ & $N_t$ \\
			\colrule
			$10.900$ & $0.239$ & $0.101$ & 20 & 40 \\
			$11.428$ & $0.149$ & $0.063$ & 32 & 64 \\
			$12.100$ & $0.103$ & $0.044$ & 48 & 96
		\end{tabular}
	\end{ruledtabular}
\end{table}

When simulating at fine lattice spacings, one must be careful to avoid any topological locking, which is known to worsen as both $\beta$ \textit{and} $N$ increase. As such, to ensure our finest lattice is not afflicted by any nonergodicity in the Markov chain, we instead choose to thermalize 100 independent hot starts with 10,000 thermalization sweeps employed for each. We can therefore be confident our configurations are representative of the ensemble.

\section{Visualizations} \label{sec:visualizations}
Our first point of consideration is the qualitative structure of $\mathrm{SU}(4)$ center-vortex matter. This is examined by extending visualization techniques previously established in Ref.~\cite{Biddle:2019gke}. To construct a three-dimensional visualization, we slice through a given dimension of the four-dimensional lattice by holding the associated coordinate fixed. This leaves one orthogonal direction in the three-dimensional slice available to identify each nontrivial plaquette in the center-projected field. A vortex segment is then rendered as an arrow existing on the dual lattice and piercing the nontrivial plaquette.

\begin{figure}
	\centering
	\includegraphics[width=\linewidth]{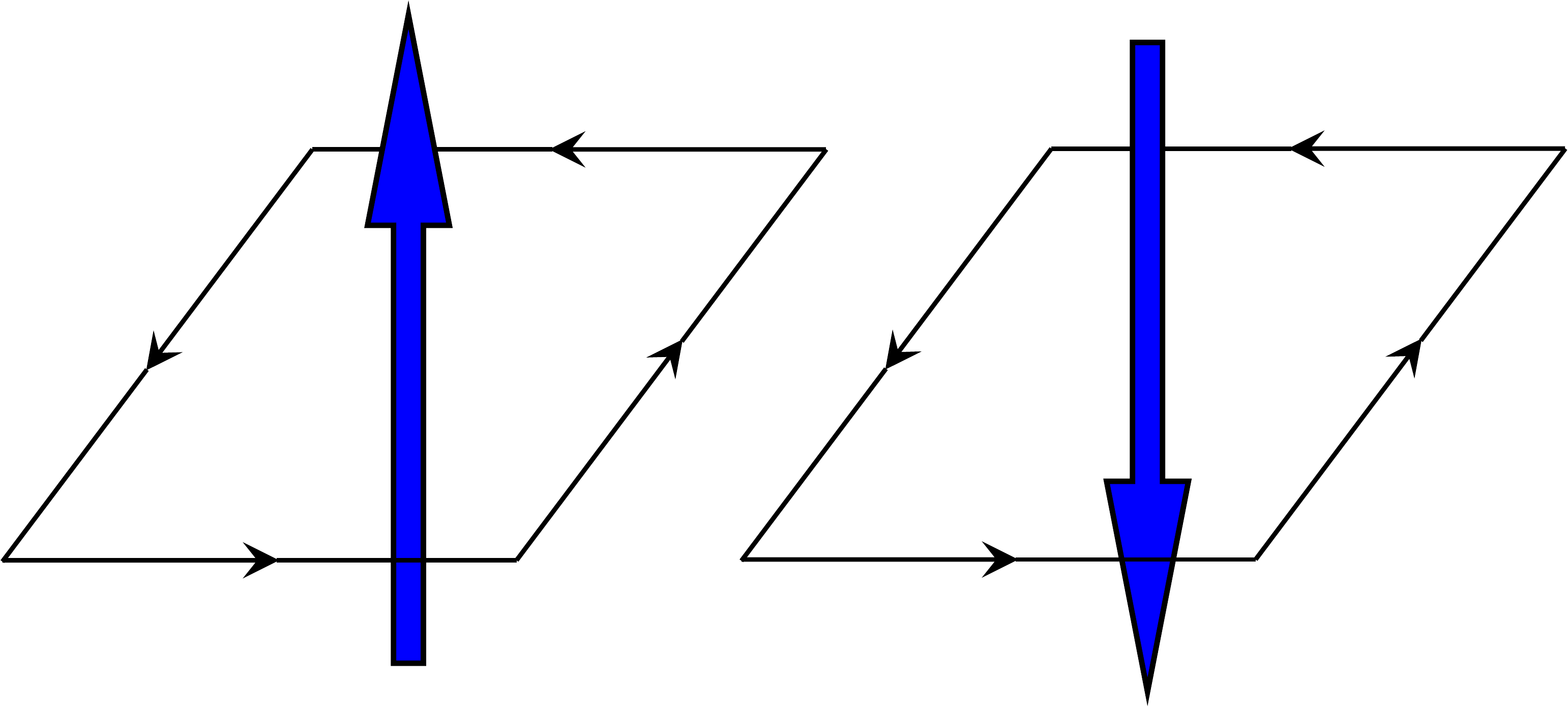}
	\vspace{-2em}
	\caption{\label{fig:SU3visconvention} The typical visualization convention for center vortices in $\mathrm{SU}(3)$. An $m = +1$ vortex segment (\textbf{left}) is represented by a jet in the available orthogonal dimension, with the direction given by the right-hand rule. An $m = -1$ vortex segment  (\textbf{right}) is rendered by a jet in the opposite direction.}
\end{figure}

Prior work has visualized exclusively $\mathrm{SU}(3)$ center-vortex structures \cite{Biddle:2019gke, Biddle:2023lod, Mickley:2024zyg, Mickley:2024vkm}. In this case, there are two distinct nontrivial center elements corresponding to $m=\pm 1$. The orientation of an $m = +1$ vortex is determined by applying the right-hand rule. Since the flow of $m = -1$ center charge is indistinguishable from an opposite flow of $m = +1$ center charge, we display an $m = -1$ vortex as a jet pointing in the opposite direction from the right-hand rule. In other words, the visualizations exclusively show the flow of $m = +1$ center charge. This convention is demonstrated in Fig.~\ref{fig:SU3visconvention}.

\begin{figure}
	\centering
	\includegraphics[width=\linewidth]{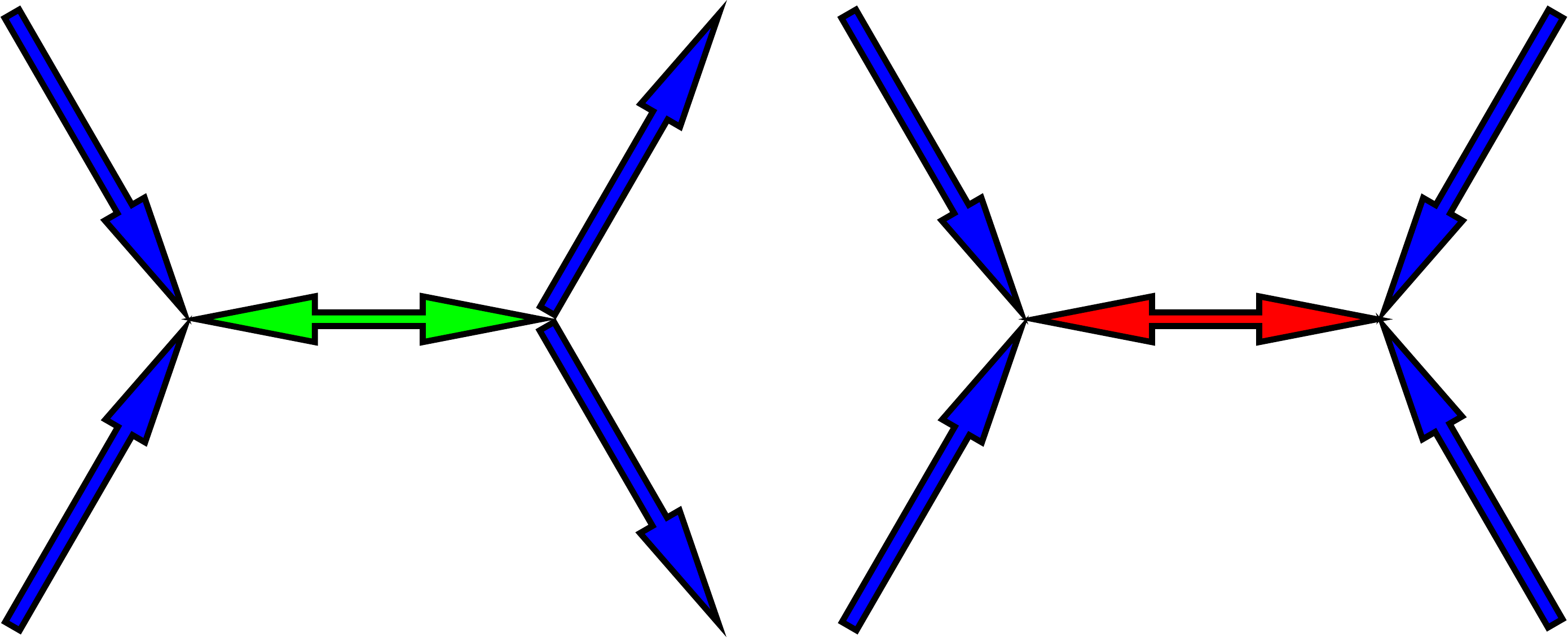}
	\vspace{-1em}
	\caption{\label{fig:SU4visconvention} Basic examples of doubly charged vortex chains in $\mathrm{SU}(4)$. A vortex convergence chain (\textbf{left}) sees two elementary vortices converge at one end of the chain and emerge at the other. A vortex matching chain (\textbf{right}) features both elementary pairs converging, or both emerging, at their respective ends of the chain. The double-sided arrow representing the doubly charged chain embodies the equivalence between a vortex charge of $m = +2$ and $m = -2$.}
\end{figure}

This process is complicated in $\mathrm{SU}(4)$ by the addition of physically inequivalent doubly charged ($m = +2$) vortices, for which there are several distinct scenarios that must be accounted for to build a comprehensive picture of $\mathrm{SU}(4)$ vortex geometry. Basic examples for two of these possibilities are illustrated in Fig.~\ref{fig:SU4visconvention}.
In the first one, two elementary vortices converge to form a single doubly charged chain, from which two elementary vortices emerge at the other end. We label this a ``vortex convergence" (VC) chain. In the second case, both pairs of elementary vortices converge at their respective ends of the chain, which we call a ``vortex matching" (VM) chain. This is allowed due to the equivalence between $m = +2$ and $m = -2$ center charges. A VM chain could also feature the pairs of elementary vortices emerging at the chain's ends, instead of converging. Naturally, the third possibility not shown in Fig.~\ref{fig:SU4visconvention} is that the doubly charged chain does not connect to any elementary vortices, forming an isolated ``loop of charge 2" (L2).

We will now describe a numerical algorithm to unambiguously identify a doubly charged vortex chain as one of these three classifications.

\begin{figure}
	\centering
	\includegraphics[height=0.55\linewidth]{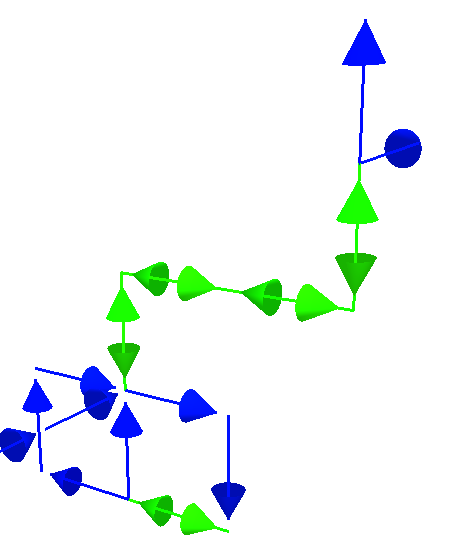}
	\hfill
	\includegraphics[height=0.55\linewidth]{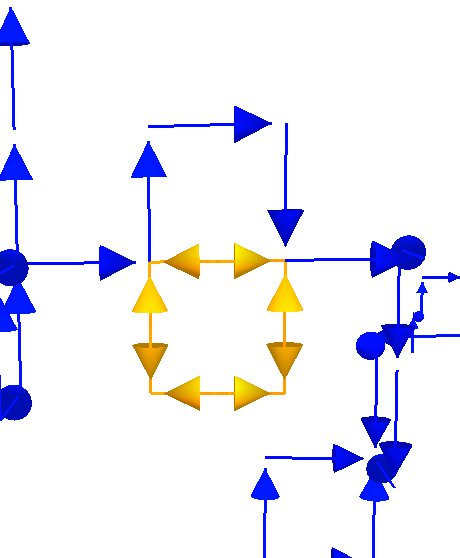}
	\caption{\label{fig:edgecases} Rare edge cases that must be accounted for in the classification of doubly charged vortex chains. Four elementary vortex segments can touch one end of the chain (\textbf{left}), with two of these attributed to forming the doubly charged chain and the remaining two constituting part of an elementary vortex line. Similarly, an L2 chain can feature points at which an elementary vortex line touches the chain (\textbf{right}).}
\end{figure}

\begin{enumerate}
	\item Select a random doubly charged vortex segment and mark it as visited. Initialize a zero-size array \texttt{touch} for later use.
	\item Identify all unvisited doubly charged vortex segments that touch the selected segment. Recursively visit each of these in turn, until no more unvisited doubly charged segments can be found. At this point, we must be at an end of the chain.
	\item Count the number of elementary vortices that touch one end of the doubly charged vortex line. For each, record if it is the tip or base of the elementary vortex that touches.
	\item Append the difference $n_\mathrm{tip} - n_\mathrm{base}$ to \texttt{touch} \textit{if} it is nonzero.
	\item Repeat Steps 3 and 4 for the other end of the doubly charged vortex.
	\item From the information contained in the \texttt{touch} array, the doubly charged chain is classified according to the following logic:
	\begin{enumerate}
		\item If \texttt{touch} is still zero-size, then it is a loop, i.e. an L2 chain.
		\item Else, if all elements of \texttt{touch} have the same sign, it is a VM chain.
		\item Else, it is a VC chain.
	\end{enumerate}
	\item Restart from Step 1 until every doubly charged vortex has been visited. This results in the classification of all such chains.
\end{enumerate}
Analyzing the two ends of the doubly charged vortex separately (in Steps 3--5) is crucial to allow for chains of length one, in which case the elementary vortices at both ends of the chain touch a single vortex segment. Furthermore, the process of calculating the difference $n_\mathrm{tip} - n_\mathrm{base}$ and considering this only when nonzero is necessary to account for several edge cases that would otherwise be incorrectly classified. Examples of these are provided in Fig.~\ref{fig:edgecases}.

The left figure shows a doubly charged chain with four elementary vortex segments touching at one end. Three of these are incoming and one is outgoing. As such, this can be resolved as two elementary vortices converging to form the doubly charged chain and a continuous elementary vortex line that happens to touch the same point. Upon examining the other end, the chain is unambiguously resolved to be VC, with the mismatch in incoming and outgoing elementary vortices accounted for by the difference $n_\mathrm{tip} - n_\mathrm{base}$, as described.

On the right, an elementary vortex line touches an L2 chain at two points. Given that the elementary vortices never combine to form the doubly charged chain, we can be assured it is L2 rather than another type. This will have $n_\mathrm{tip} - n_\mathrm{base} = 0$ at both touching points and as such will be disregarded, leading to the correct classification.

One hypothetical scenario not accounted for by the above prescription is the intersection of two doubly charged chains. According to the classification procedure, this would be identified as a single chain. In principle, there would be ambiguity in how the intersection is resolved into two chains. However, no such instance was encountered in our configurations, meaning we can be confident in the outlined algorithm.

With that covered, the rendering color scheme for the visualizations is as follows:\ VC chains are in green, VM chains are in red, and L2 chains are in yellow/orange. Elementary vortices that are part of the percolating cluster will be colored dark blue, or light blue if in a secondary cluster.

\begin{figure*}
	\centering
	\includegraphics[width=0.88\linewidth]{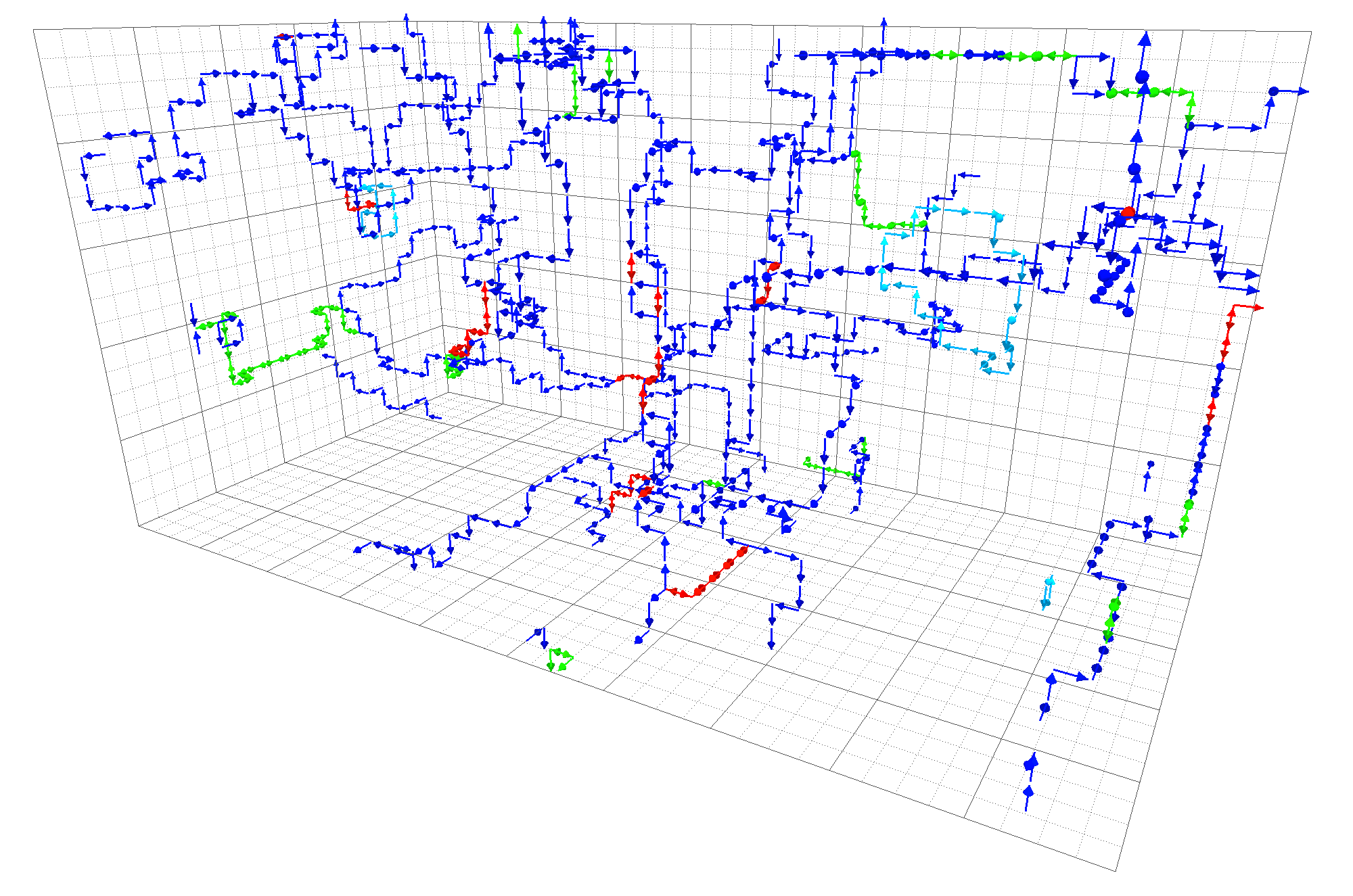}
	\vspace{-1em}
	\caption{\label{fig:coarse_vis} $\mathrm{SU}(4)$ center-vortex structure in spatial slices at $\beta = 10.900$, corresponding to $a\sqrt{\sigma} \simeq 0.239$. The three-dimensional volume is $20 \times 20 \times 40$. Single arrows illustrate the direct flow of charge $+1$ vortices. Double-arrow vortex convergence (green, light gray) and vortex matching (red, dark gray) chains are found throughout the percolating cluster. Several elementary-vortex secondary clusters (light blue, light gray) are also present in the volume.}
\end{figure*}

\begin{figure*}
	\centering
	\includegraphics[width=0.88\linewidth]{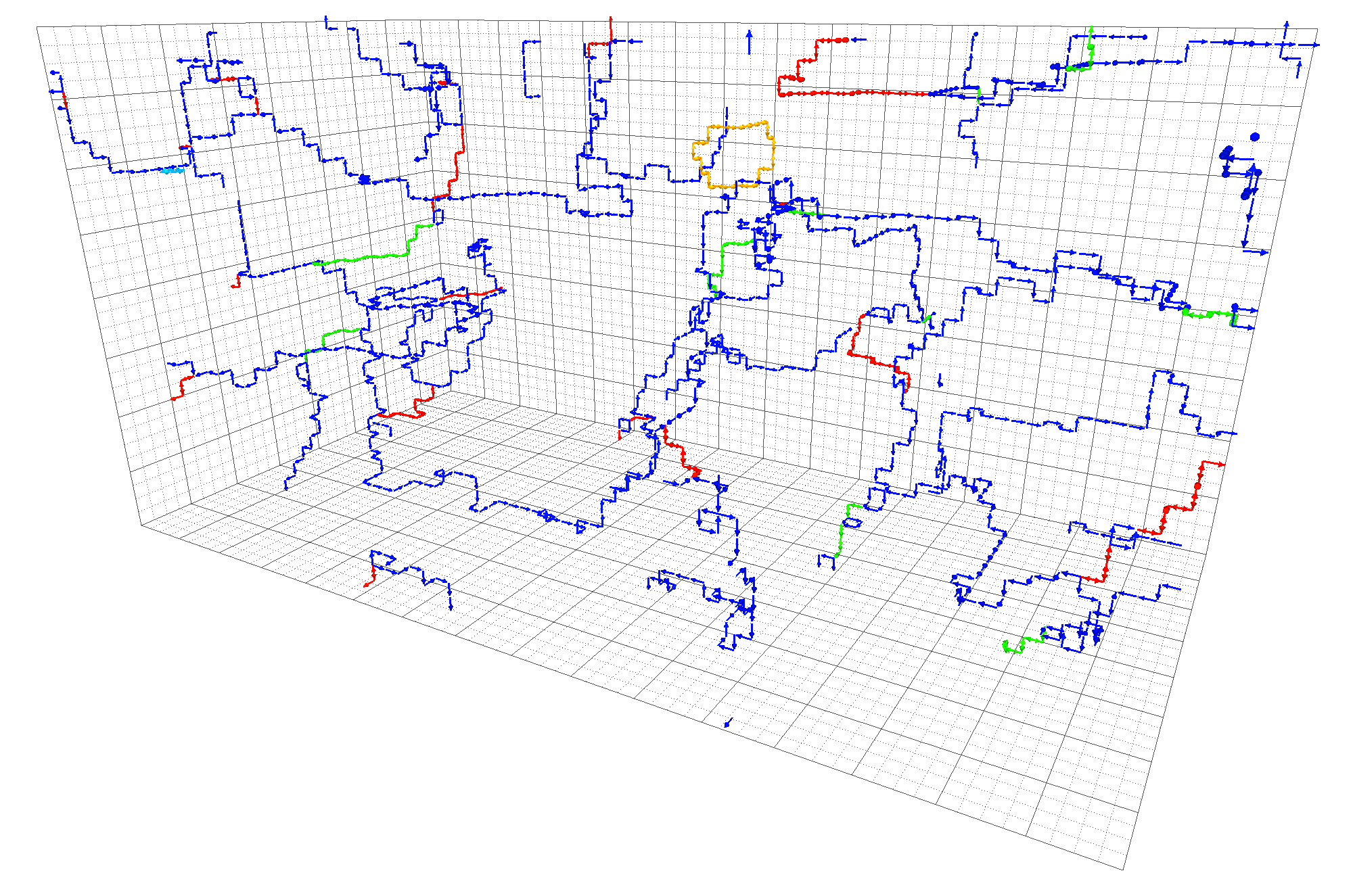}
	\vspace{-1em}
	\caption{\label{fig:medium_vis} $\mathrm{SU}(4)$ center-vortex structure in spatial slices at $\beta = 11.428$, corresponding to $a\sqrt{\sigma} \simeq 0.149$. The three-dimensional volume is $32 \times 32 \times 64$. Illustrated as in Fig.~\ref{fig:coarse_vis}. An L2 chain can be found in the upper middle of this slice.}
\end{figure*}

\begin{figure*}
	\centering
	\vspace{0.5em}
	\includegraphics[width=0.88\linewidth]{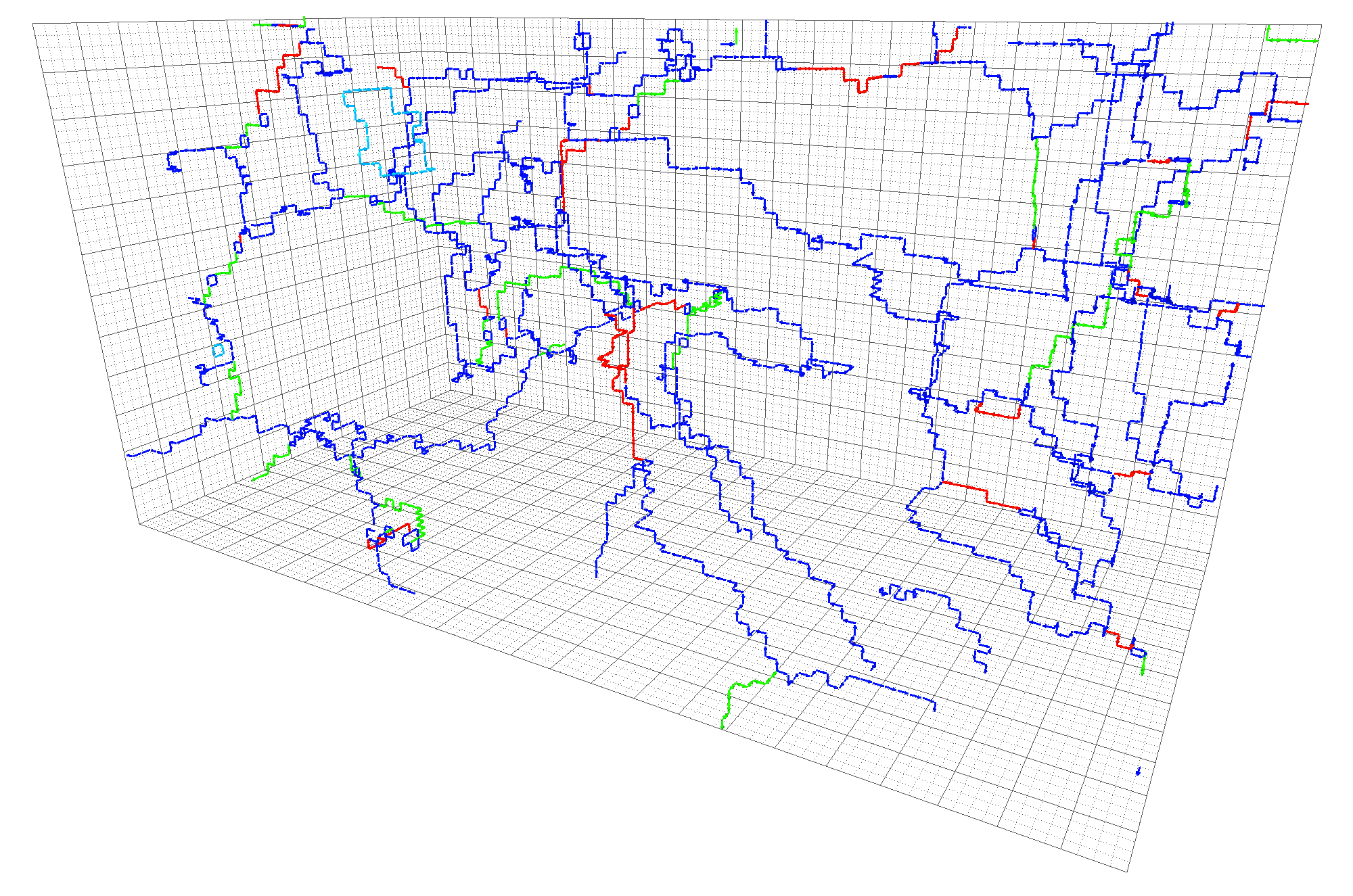}
	\vspace{-1em}
	\caption{\label{fig:fine_vis} $\mathrm{SU}(4)$ center-vortex structure in spatial slices at $\beta = 12.100$, corresponding to $a\sqrt{\sigma} \simeq 0.103$. The three-dimensional volume is $48 \times 48 \times 96$. Illustrated as in Fig.~\ref{fig:coarse_vis}. Here we see the persistence of both VC and VM doubly charged chains as the lattice spacing is decreased.}
\end{figure*}

We now present visualizations for each lattice spacing. Given the vortex sheet percolates all four dimensions at low temperatures, the observed three-dimensional structure is insensitive to the choice of dimension sliced over. We choose to show ``spatial slices," obtained from fixing one of the three spatial ($x$-$y$-$z$) coordinates. This is to take advantage of the larger associated three-dimensional volume, which incorporates the long time dimension. A typical spatial slice is displayed for each lattice spacing throughout Figs.~\ref{fig:coarse_vis}--\ref{fig:fine_vis}.

The visualizations reveal a percolating cluster that dominates the structure, with only a handful of smaller secondary clusters scattered throughout the lattice. This is in line with prior $\mathrm{SU}(3)$ visualizations at low temperatures \cite{Biddle:2019gke, Mickley:2024zyg}. The presence of doubly charged chains connecting the elementary-vortex components is manifest. Their total length in the network is seen to be significantly smaller than that of the elementary vortices, and at a glance there appears to be approximately equal proportions of vortex convergence and vortex matching chains. This will be investigated in detail in Sec.~\ref{subsec:proportions}.

Before producing the visualizations, it was unclear whether the doubly charged vortex chains, in particular vortex convergences, would survive in the continuum limit. One could imagine that what appears as an abundance of vortex coalescence outside the continuum limit gives place to a network only formed by elementary vortex lines in a finer lattice. Our finest lattice visualized in Fig.~\ref{fig:fine_vis} would suggest this is not the case, still featuring an abundance of VC chains. In particular, this implies they are physical objects.

\begin{figure}
	\centering
	\fbox{\includegraphics[width=0.96\linewidth]{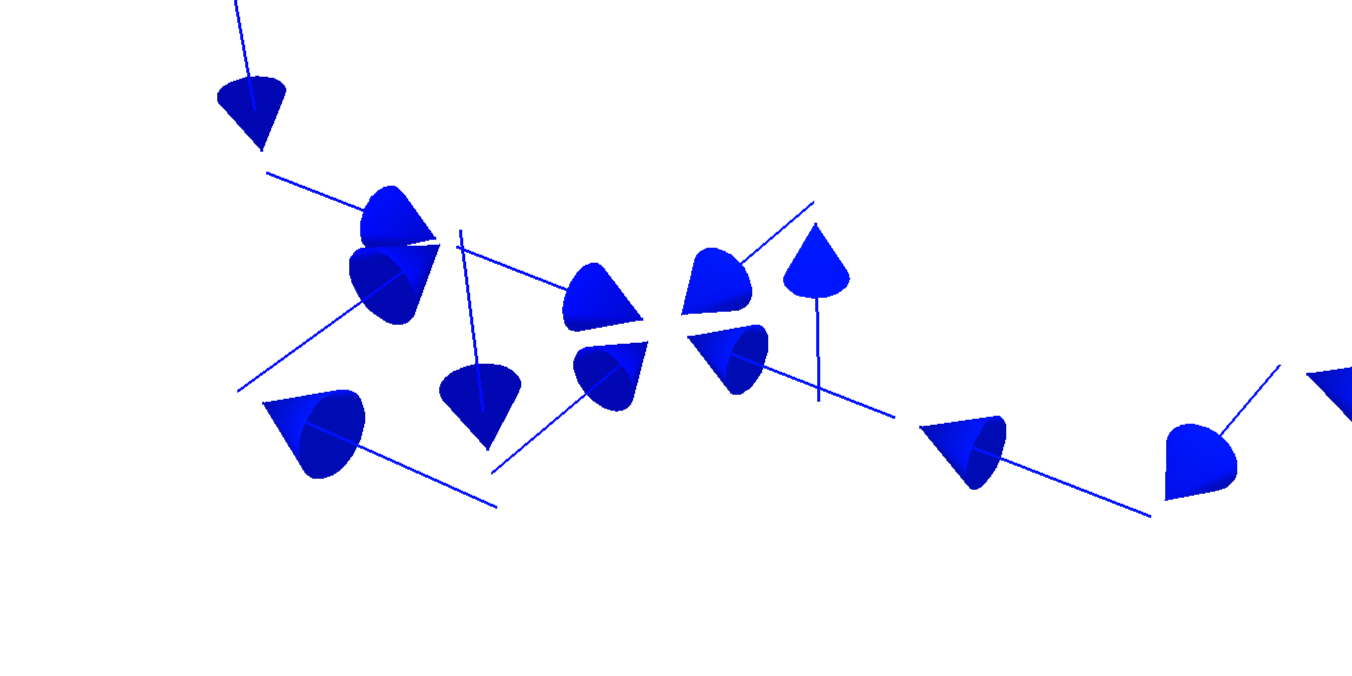}}
	
	\vspace{-0.15em}
	
	\fbox{\includegraphics[width=0.96\linewidth]{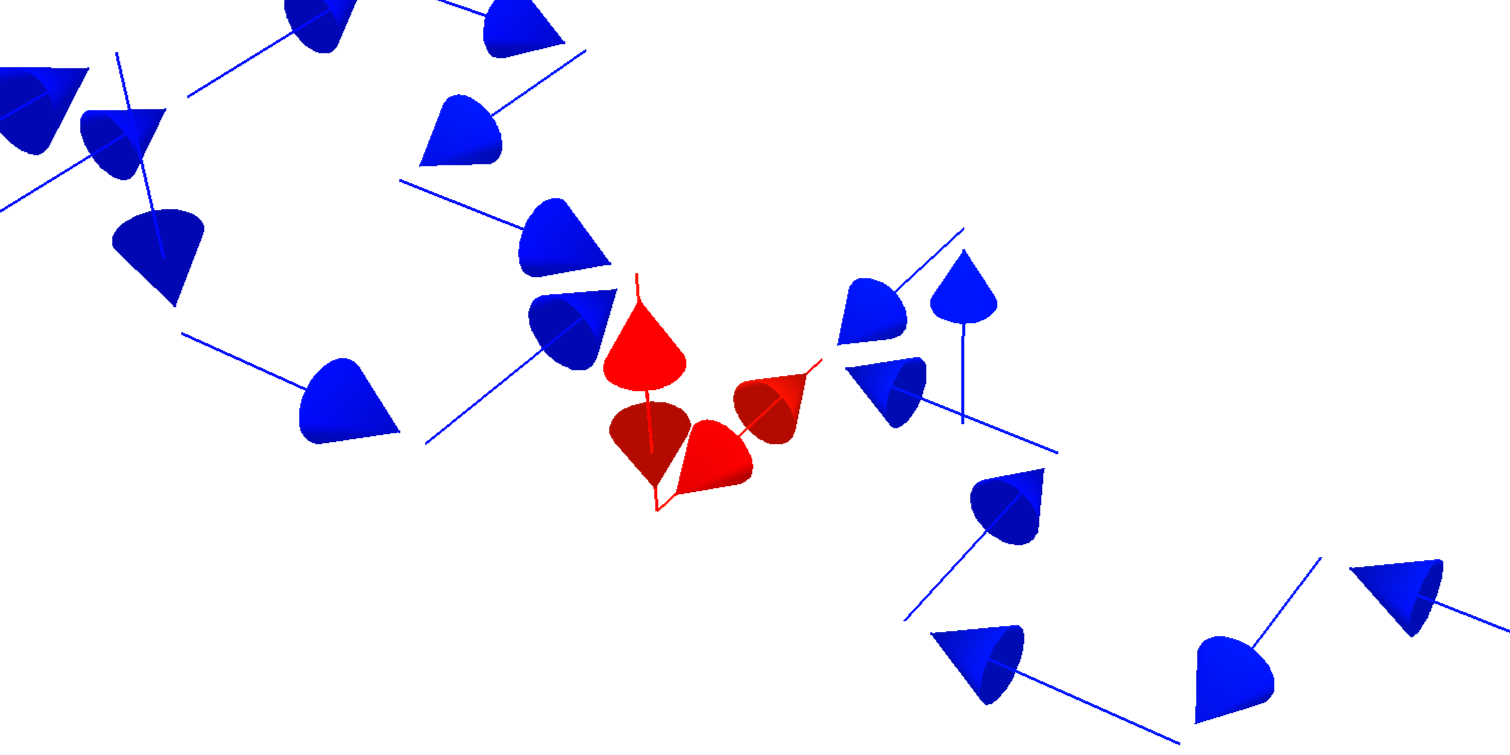}}
	
	\vspace{-0.15em}
	
	\fbox{\includegraphics[width=0.96\linewidth]{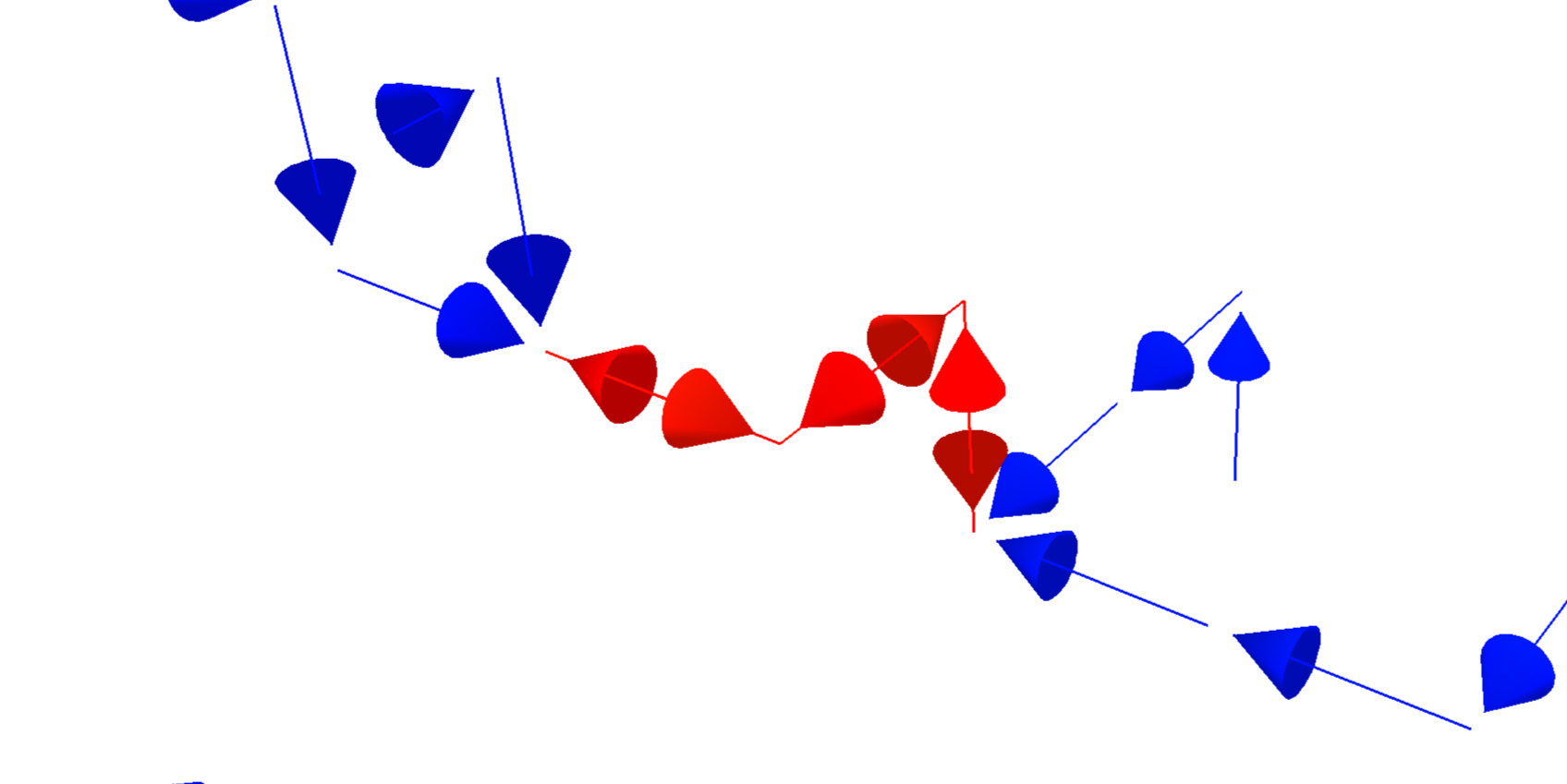}}
	\caption{\label{fig:extended_monopole} Three consecutive slices from our medium lattice ($\beta = 11.428$) showing the evolution of vortex structure. We see a four-way center monopole (\textbf{top}) giving rise to an intermediate vortex matching chain (\textbf{middle}/\textbf{bottom}). VM chains can therefore be viewed as ``extended center monopoles."}
\end{figure}

The visualizations also provide an alternative interpretation of vortex matching chains demonstrated in Fig.~\ref{fig:extended_monopole}, which shows a zoom of three consecutive slices from our medium lattice spacing on a particular region.
The first frame reveals a four-way center monopole in which four elementary vortices are matched at a point, allowed due to the conservation of center charge modulo $N$ in $\mathrm{SU}(N)$ Yang-Mills theory. These are analogs of the three-way center monopoles/branching points that exist in $\mathrm{SU}(3)$. The subsequent two frames show the center monopole giving rise to an intermediate VM chain. This elicits VM chains as ``extended center monopoles," in the sense that they can always be thought of as a monopole where the four emanating lines coalesce in pairs over some extent.

Interactive graphics of the slices displayed in Figs.~\ref{fig:coarse_vis}--\ref{fig:fine_vis} are provided in the supplemental material throughout Figs.~\ref{fig:coarse_supp}--\ref{fig:fine_supp}. We invite the reader to explore these models and the numerous examples of each type of doubly charged object that they provide.

In addition, the supplementary animations reveal the complex four-dimensional nature of the vortex sheet. It is instructive to compare the animations corresponding to our coarsest and finest lattices, of which the latter appears considerably smoother owing to the smaller physical distance between consecutive frames/slices. To an extent, this allows the doubly charged chains to be tracked over the animation. Careful inspection of the two finer animations (Figs.~\ref{fig:medium_supp} and \ref{fig:fine_supp}) also reveals how some secondary clusters separate from and merge with the percolating cluster. This indicates that they lie in the same connected surface in four dimensions, only appearing as disconnected loops in three-dimensional slices due to the surface's curvature. This idea was also discussed and illustrated in Ref.~\cite{Mickley:2024vkm}.
	
\section{Vortex chain analysis} \label{sec:analysis}
In this section, we analyze a selection of vortex statistics to quantify the behavior qualitatively identified through the visualizations. Our first focus is on the various types of doubly charged chains. We start by considering the proportions of doubly charged vortices, both in comparison to elementary vortices and the relative proportions among the different types of chains. Thereafter, we utilize the chain-identification algorithm described in Sec.~\ref{sec:visualizations} to analyze the lengths of doubly charged chains and how these scale with the lattice spacing. Initially, their average lengths are studied in Sec.~\ref{subsec:lengths}, before a deep dive into the inherent distribution of lengths is undertaken in Sec.~\ref{subsec:probabilities}.

All statistical quantities are computed with 100 bootstrap ensembles, with errors calculated as the standard deviation of the bootstrap estimates.
	
\subsection{Proportions} \label{subsec:proportions}
We start by investigating the proportions of total lengths occupied by elementary and doubly charged vortices in the network. These are calculated in terms of the number of dual links visited in each case, or simply as
\begin{equation}
	p_{1(2)} = \frac{\text{\# of plaquettes with } m = \pm 1 \text{ } (m = 2)}{\text{\# of pierced plaquettes}} \,.
\end{equation}
The values provided in Table~\ref{tab:proportions} quantitatively verify that the vortex matter is dominated by elementary vortices ($m = \pm 1$), and that doubly charged vortices contribute only $\approx 10$--$15$\% of all pierced plaquettes. The proportion $p_1$ is equally split between vortices with $m = +1$ and $m = -1$. This is to be anticipated as there is no preferred orientation for the flow of center charge.
\begin{table}
	\centering
	\caption{\label{tab:proportions} The various proportions discussed in text, including the total length proportion of elementary and doubly charged vortices ($p_1$ and $p_2$, respectively) and the relative proportions among the various doubly charged chain types ($p_\mathrm{VC}$, $p_\mathrm{VM}$ and $p_\mathrm{L2}$).}
	\begin{ruledtabular}
		\begin{tabular}{c|cc|ccS[table-format=1.5]}
			$a \, (\sigma^{-1/2})$ & $p_1$ & $p_2$ & $p_\mathrm{VC}$ & $p_\mathrm{VM}$ & \multicolumn{1}{c}{$p_\mathrm{L2} \, (\times 10^{-2})$} \\
			\colrule
			$0.239$ & 0.873(1) & 0.127(1) & 0.461(2) & 0.537(2) & 0.159(11) \\
			$0.149$ & 0.867(1) & 0.133(1) & 0.460(1) & 0.539(1) & 0.067(4) \\
			$0.103$ & 0.856(1) & 0.144(1) & 0.466(1) & 0.533(1) & 0.042(2)
		\end{tabular}
	\end{ruledtabular}
\end{table}

It is interesting to observe a soft shift in the relative proportions of elementary and doubly charged vortices as the lattice spacing is reduced, in favor of the doubly charged portion. This is perhaps counterintuitive under the assumption that doubly charged vortices could be separated into two elementary vortices with a finer resolution, which would effect the opposite shift. Therefore, this finding corroborates that doubly charged vortices are physical objects that persist in the continuum limit.

In addition, despite the mild increase in proportion of doubly charged objects on the finer lattice, it seems reasonable to claim that elementary and doubly charged vortices remain inequivalent in the continuum with a strong preference for elementary vortices. This is intuitive as the creation of a doubly charged vortex requires additional action compared to an elementary vortex (as governed by the real part of the plaquette), leading to a suppression of the former relative to the latter.

It is also instructive to compare relative proportions amongst the various types of doubly charged chains identified by the algorithm in Sec.~\ref{sec:visualizations}. That is, of all such chains, we compute the fraction that are vortex convergence ($p_\mathrm{VC}$), vortex matching ($p_\mathrm{VM}$) or an isolated doubly charged loop ($p_\mathrm{L2}$). These are also provided in Table~\ref{tab:proportions}. Clearly, L2 chains contribute a negligible amount towards the quantity of doubly charged vortex matter.

The findings reveal a slight tendency for VM chains to be formed over VC, with an approximately 54:46 split. This discrepancy is slightly smaller on the finer lattice, though still statistically highly significant. The origin for this preference can be understood as due to the different robustness properties implied by center charge conservation, which must be satisfied locally. Consider four elementary vortex lines crossing a given closed surface, and suppose this persists over a sequence of time slices. If all the vortices enter the enclosed volume $\mathcal{V}$, then they will remain correlated along the slices. The flux along a single vortex line cannot change from $+1$ to $-1$. That is, inside $\mathcal{V}$, the four lines will always be connected with the presence of at least a VM chain or a center monopole, which can nucleate a VM chain in a subsequent slice but cannot split into two parts. This would suggest that VM objects are persistent, unable to be separated due to a ``tie" inside the chain that binds the pairs of elementary vortices together. Thus, VM objects can disappear, but through a process involving a four-way center monopole or a VM chain with opposite orientation.

On the other hand, when two elementary vortices enter $\mathcal{V}$ and two vortices leave, they are not necessarily correlated inside. If a vortex convergence chain is formed, it can disappear more easily by simply bifurcating into the pair of elementary vortex lines that formed the VC chain. Hence, after taking into account all three-dimensional slices, one would expect a modest preference for VM objects owing to their relative persistence across slices.

\subsection{Chain lengths} \label{subsec:lengths}
Another measure made available through the chain-identification algorithm is the lengths of the individual chains, which can be deduced by keeping track of the number of doubly charged dual links visited in each iteration of the algorithm. In this section, we will consider simply the average lengths of each doubly charged chain type. These values are provided both in lattice and physical units in Table~\ref{tab:lengths}.
\begin{table}
	\centering
	\caption{\label{tab:lengths} The average chain lengths for each type of doubly charged chain, in both lattice units (indicated by the hat) and in physical units (expressed in terms of the string tension). The average L2 chain length is dominated by $1 \times 1$ loops, and as such is provided only in lattice units. VM chains are seen to be on average shorter than VC, and both decrease slightly with the lattice spacing. The continuum ($a = 0$) values, obtained via a linear extrapolation with $a^2$ as described in text, are also provided. The discrepancy between VC and VM chains is seen to persist in the continuum limit.}
	\begin{ruledtabular}
		\begin{tabular}{c|ccS[table-format=1.5]|cc}
			$a \, (\sigma^{-1/2})$ & $\hat{\ell}_\mathrm{VC}$ & $\hat{\ell}_\mathrm{VM}$ & \multicolumn{1}{c|}{$\hat{\ell}_\mathrm{L2}$} & $\ell_\mathrm{VC} \, (\sigma^{-1/2})$ & $\ell_\mathrm{VM} \, (\sigma^{-1/2})$ \\
			\colrule
			$0.239$ & 3.69(2) & 3.40(2) & 4.26(5)  & 0.881(5) & 0.811(4) \\
			$0.149$ & 5.42(3) & 4.97(3) & 4.53(16) & 0.809(5) & 0.741(5) \\
			$0.103$ & 7.54(4) & 7.08(4) & 4.71(22) & 0.775(4) & 0.728(4) \\
			\colrule
			$0$ & -- & -- & \multicolumn{1}{c|}{--} & 0.754(4) & 0.706(4)
		\end{tabular}
	\end{ruledtabular}
\end{table}

Looking first at secondary loops of charge 2, we find the average lengths for each spacing is slightly above four lattice units. This is primarily determined by the fact that most L2 chains are simple $1 \times 1$ doubly charged vortex loops visiting four dual links (such as seen in Fig.~\ref{fig:edgecases}), the minimum amount to form a closed loop. The average length does increase slightly as the lattice spacing is reduced, indicating the presence of larger L2 chains that would be comparable in \textit{physical} units across the three lattices. However, with the principal contribution being $1 \times 1$ loops in each case, discretization effects are likely to be dominant.

Turning to vortex convergence and matching chains, we find that VM objects are on average a little shorter than VC across all lattice spacings. That is, VM chains tend to be shorter but are also more prevalent. Regarding the lattice-spacing dependence, one might ideally expect the physical chain lengths to remain invariant in reducing the spacing. This would imply the lengths in lattice units scale up, which is predominantly the observed behavior. Still, the physical lengths are found to exhibit a soft dependence on the lattice spacing, decreasing lightly as $a \to 0$.

\begin{figure}
	\centering
	\includegraphics[width=\linewidth]{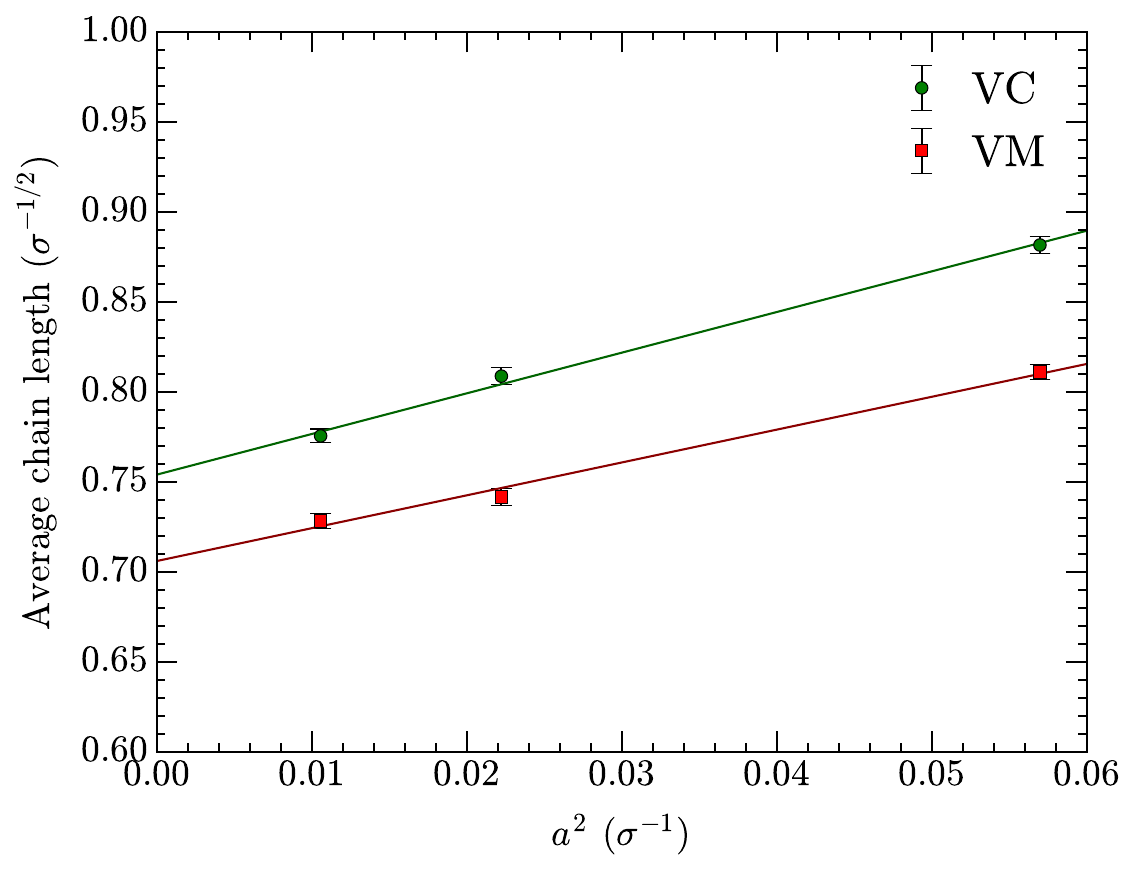}
	
	\vspace{-0.5em}
	
	\caption{\label{fig:average_chain_length} The average chain lengths for VC and VM objects as a function of $a^2$, with linear extrapolations to the continuum limit. The fits are seen to capture the data trend, with reduced $\chi^2$ values of $\chi^2/\mathrm{d.o.f.} = 1.39$ for VC and $1.70$ for VM. The difference between VC and VM chain lengths is seen to persist in the continuum limit (at $a = 0$).}
\end{figure}

In reality, since the basic plaquette action used to generate the ensembles has $\order{a^2}$ errors, we would anticipate physical quantities to depend linearly on $a^2$ provided higher-order terms are negligible. This motivates plotting the physical lengths as a function of $a^2$ and investigating whether a linear extrapolation can be performed to the continuum limit. This is presented in Fig.~\ref{fig:average_chain_length}.

We find that the physical lengths are approximately linear with $a^2$, and as such fits are overlaid that allow an extrapolation to the continuum limit. These are seen to describe the data well, passing through all points within statistical uncertainty even if at the edge of the bounds. This is corroborated quantitatively by the $\chi^2/\mathrm{d.o.f.}$ values for the fits, which are $1.39$ for VC and $1.70$ for VM; these are both $\sim\mathcal{O}(1)$. In addition to statistical uncertainty, the variation around the linear trend line could also be due to next-to-leading-order errors. The values extracted at $a = 0$ are provided in Table~\ref{tab:lengths}. These extrapolations suggest the physical lengths remain finite in the continuum limit.

Looking at the extrapolated continuum values, we find that the difference between VC and VM chain lengths does persist in the continuum limit. The distinction is smaller compared against our coarsest lattice, though is still statistically strongly significant. This importantly indicates the tendency for vortex matching chains to be shorter is unlikely to be a lattice artifact. The difference in VM and VC chain lengths may be related to the reason VM chains are more prevalent than VC. As previously described, small VC chains can easily dissociate into two elementary vortices, making small VC chains rare. In contrast, short VM objects, as extended center monopoles, are more persistent. Thus, when averaging over all three-dimensional slices, this could result in the observed smaller average VM chain length.

\subsection{Splitting probabilities} \label{subsec:probabilities}
\begin{figure*}
	\includegraphics[width=0.48\linewidth]{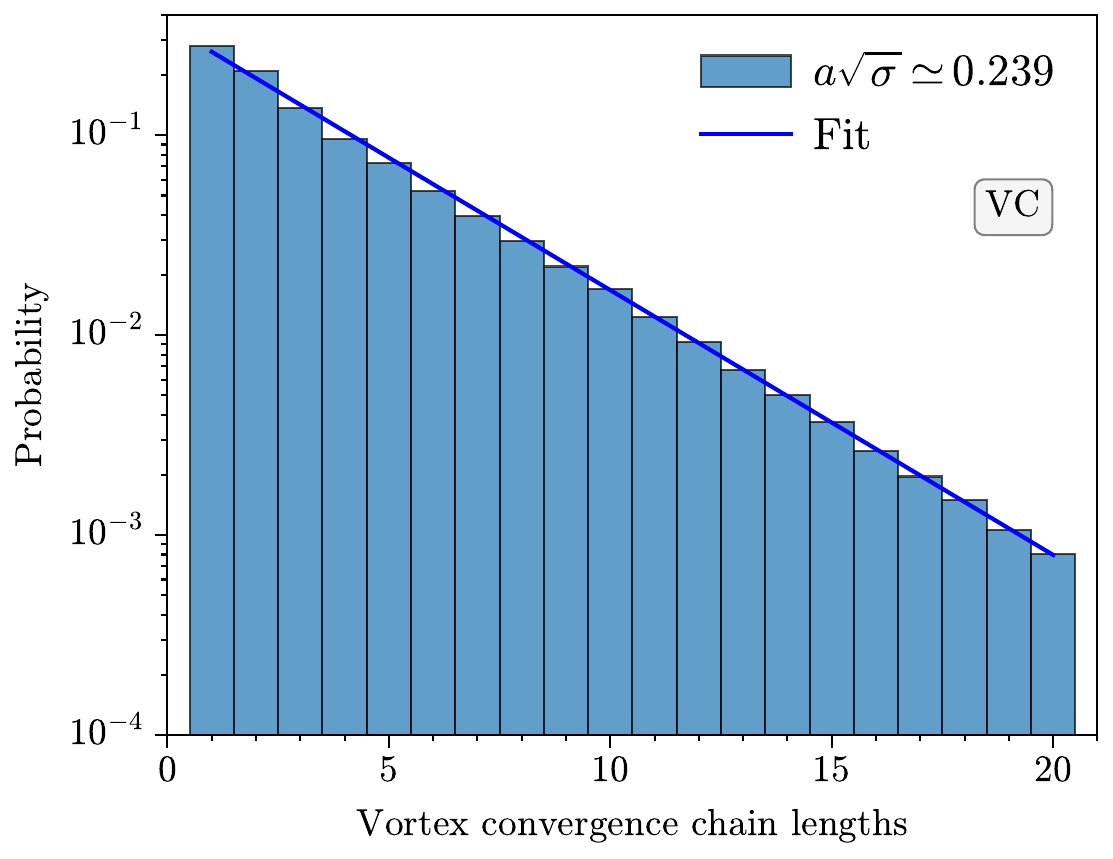}
	\hfill
	\includegraphics[width=0.48\linewidth]{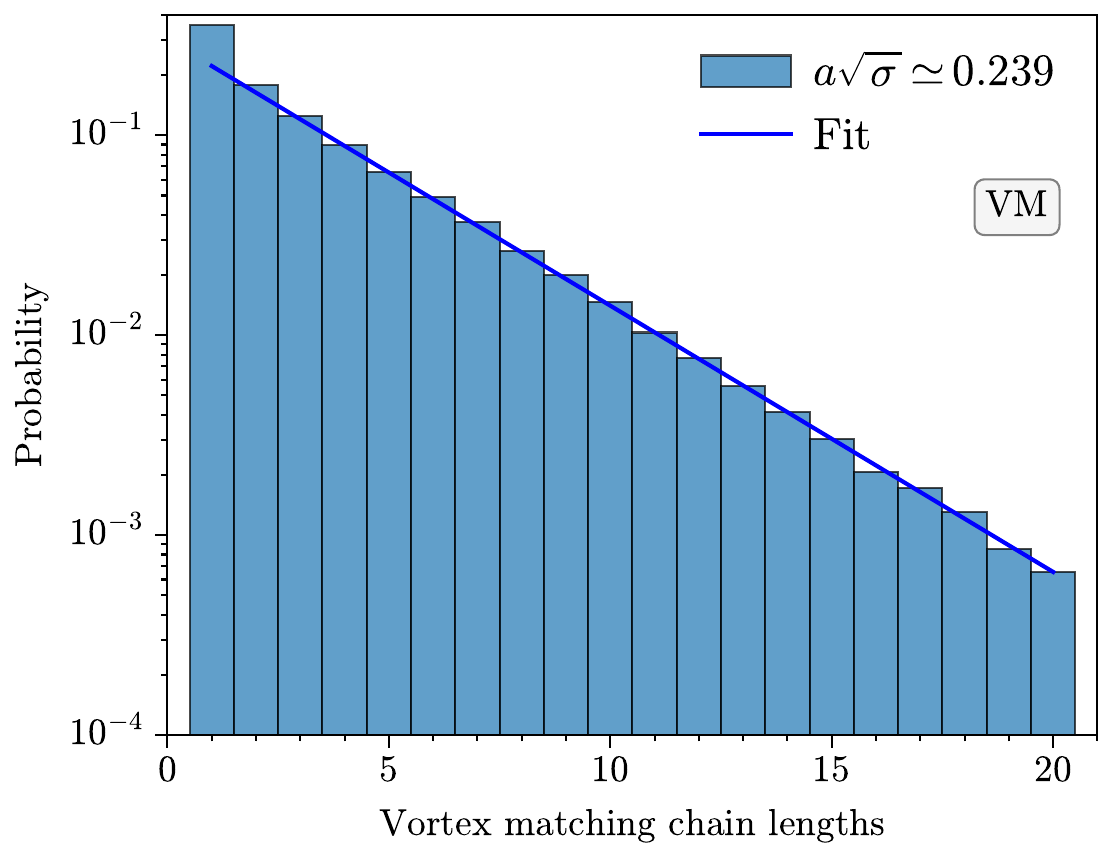}
	
	\vspace{1em}
	
	\includegraphics[width=0.48\linewidth]{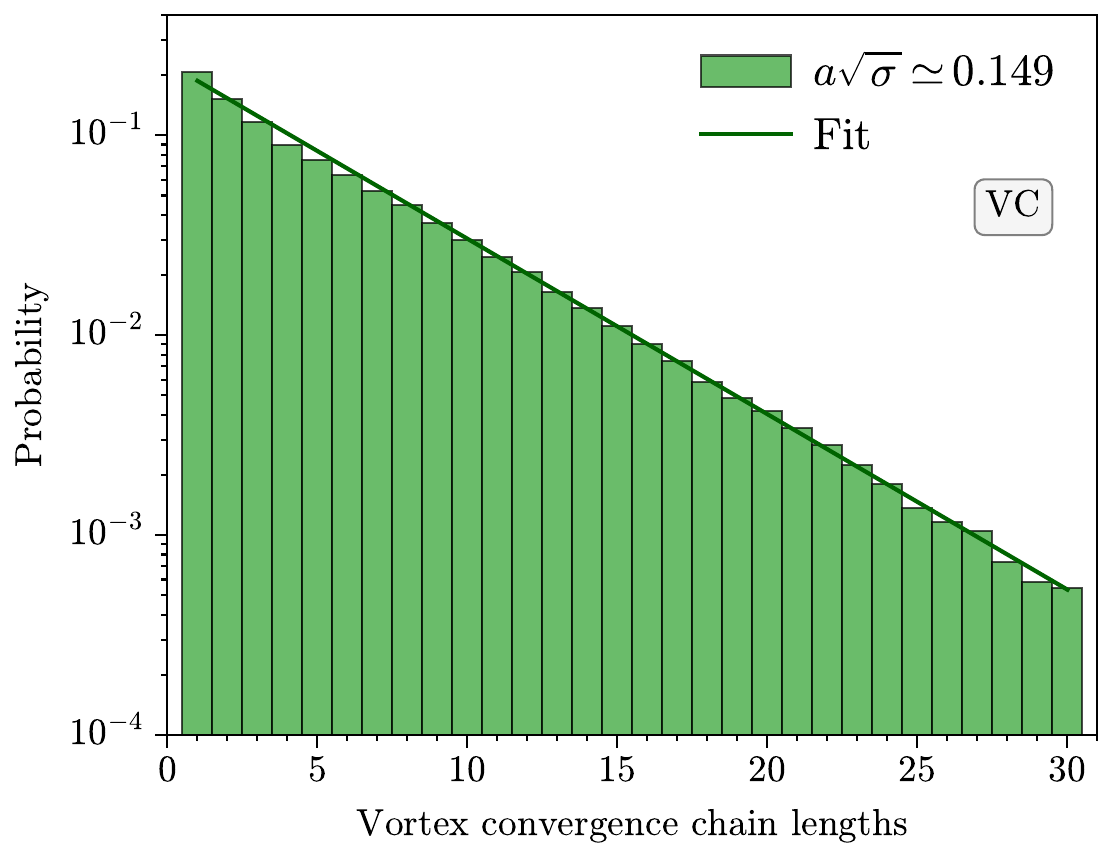}
	\hfill
	\includegraphics[width=0.48\linewidth]{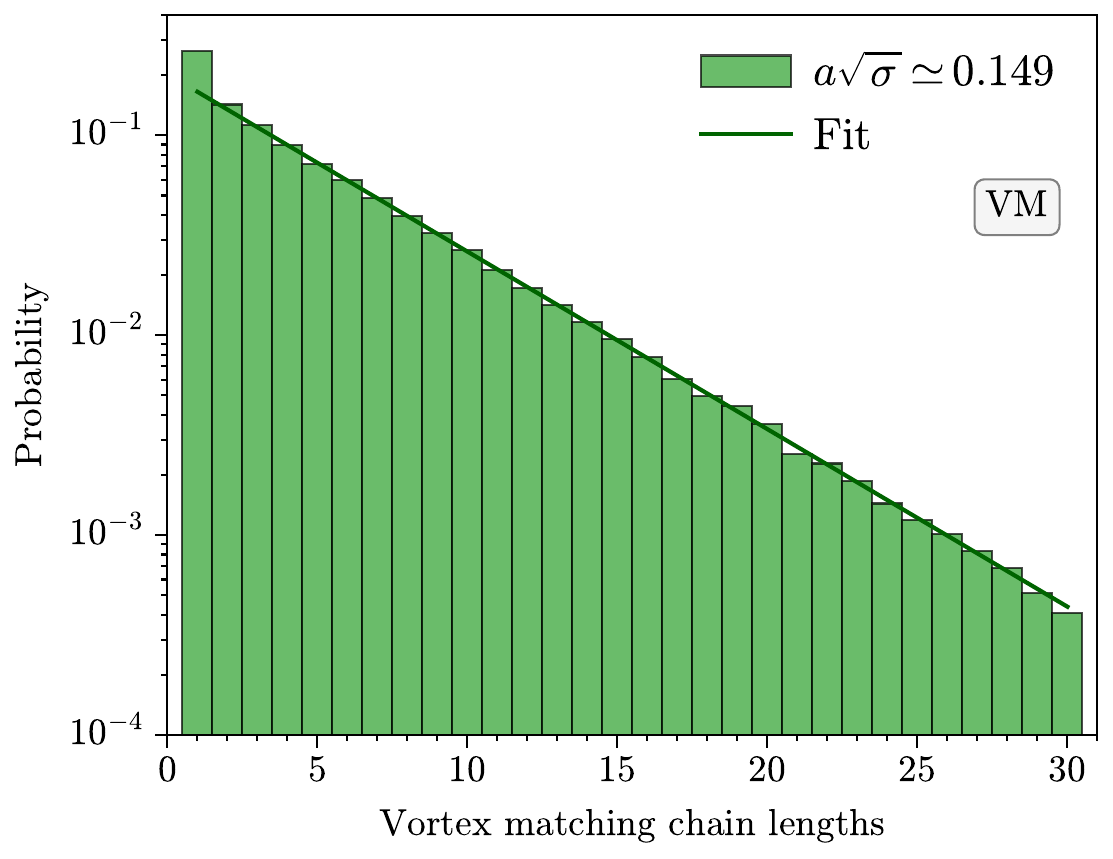}
	
	\vspace{1em}
	
	\includegraphics[width=0.48\linewidth]{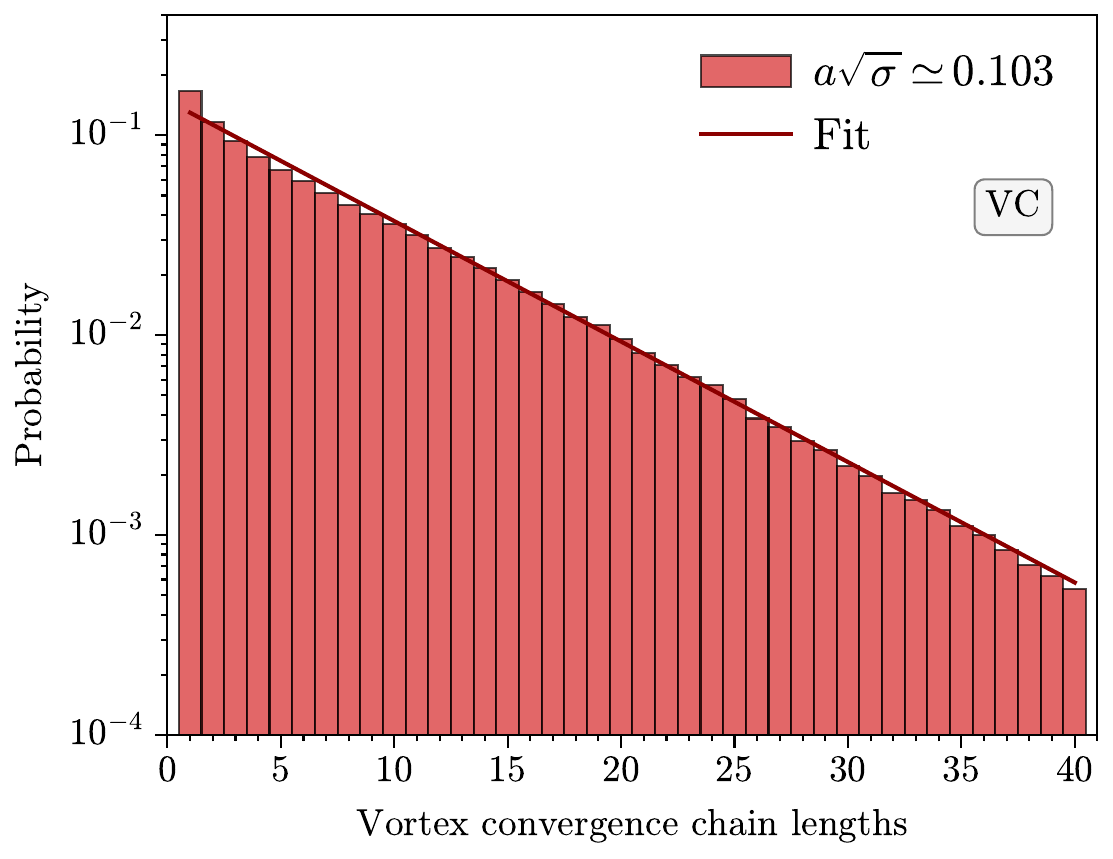}
	\hfill
	\includegraphics[width=0.48\linewidth]{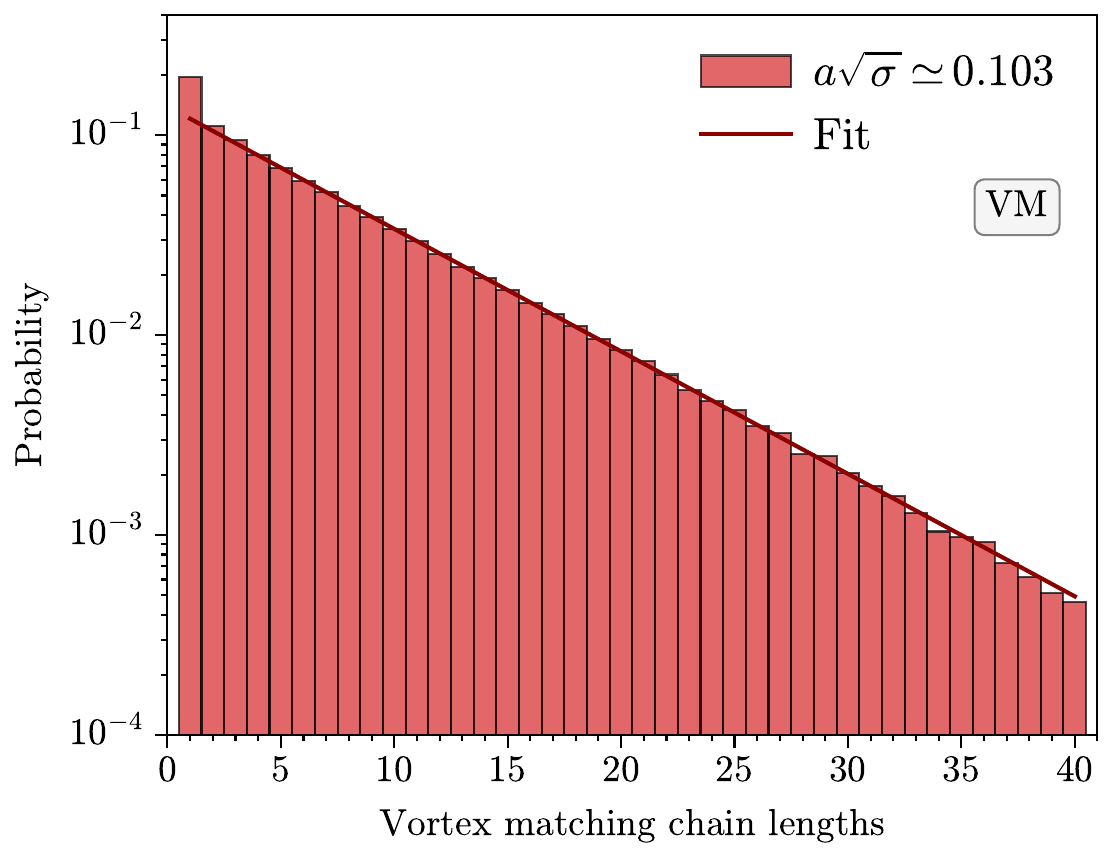}
	
	\vspace{-0.5em}
	
	\caption{\label{fig:histograms} The distributions of chain lengths on a logarithmic scale for VC (\textbf{left}) and VM (\textbf{right}) chains at each lattice spacing under consideration:\ $a\sqrt{\sigma} \simeq 0.239$ (\textbf{top}), $0.149$ (\textbf{middle}) and $0.103$ (\textbf{bottom}). All histograms are seen to be predominantly linear, implying the chain lengths are exponentially distributed. We note the exception of a length of one for VM chains, which is strongly preferred. The exponential fits for chain lengths $> 10$ described in text are also overlaid. These fits are more accurate for VM chains, revealing that the probabilities for short VC chain lengths tend to underestimate the linear trend. This is in accord with our discussion of average chain lengths and the idea that short VC chains are relatively easy to break into two elementary vortices.}
\end{figure*}

In addition to the average chain lengths, we now proceed to scrutinize the intrinsic \textit{distribution} of doubly charged chain lengths. This will reveal more subtle differences between VC and VM chains not apparent through the bulk quantities already considered. For this, we create histograms of the calculated chain lengths such that the vertical axis provides the probability for a given chain length to occur. These are produced separately for VC and VM objects, and are presented at each lattice spacing on a logarithmic scale in Fig.~\ref{fig:histograms}.

The histograms for both types of doubly charged chains are seen to be predominantly linear, which due to the logarithmic scale implies the chain lengths are exponentially distributed. The primary exception to this categorization is a pronounced preference for specifically VM chains to have a length of one. As this is applicable only to VM objects, this is certainly a factor as to why VM chains tend to be shorter than VC. That said, given Fig.~\ref{fig:histograms} shows the length in lattice units, the fact that this clustering at a length of one exists irrespective of the lattice spacing indicates it is likely a lattice artifact that will vanish in the continuum limit. This could be related to the observation that the difference in average chain lengths does diminish slightly in approaching the continuum, as seen in Fig.~\ref{fig:average_chain_length}.

The exponential distributions followed by both types of doubly charged objects are similar to an adjacent finding in $\mathrm{SU}(3)$, in which the chain lengths between branching points (the three-way center monopoles) were also observed to be exponentially distributed, barring a clustering at short separations (chain lengths $\leq 3$) \cite{Biddle:2023lod, Mickley:2024zyg}. This signified that a vortex line has a constant probability of branching as it propagates through spacetime, with this probability extracted from exponential fits to the distribution.

Analogously, a probability can be inferred from the distributions here in Fig.~\ref{fig:histograms}. This is interpreted as the probability for a doubly charged chain to ``split" back into two elementary vortices at a given point along the chain. As such, we expect this probability to have an inverse relation to the chain lengths---the greater the ``splitting probability," the shorter the chains will tend to be. To investigate this, exponential fits of the form
\begin{equation} \label{eq:exponential}
	P(n) = \zeta \, e^{-\beta n}
\end{equation}
are performed to the VC and VM chain-length distributions. Under a constant probability $q$ to split at each step along a chain, the probability to split at the $n$th step is
\begin{equation} \label{eq:geodist}
	P(n) = (1 - q)^{n-1} \, q \,,
\end{equation}
such that $q$ can be obtained from $\beta$ in the exponential fit by equating Eqs.~(\ref{eq:exponential}) and (\ref{eq:geodist}),
\begin{equation}
	P(n) = (1 - q)^{n-1} \, q = \zeta \, e^{-\beta n} \,.
\end{equation}
Taking the $\log$ and gathering the $n$-dependent terms,
\begin{equation}
	q = 1 - e^{-\beta} \,.
\end{equation}
These fits are carried out exclusively for chain lengths $n > 10$, as a close inspection of the distributions for VC chains reveals that the probabilities at short chain lengths tend to underestimate the linear trend line. This restriction is then also applied to VM chains for consistency. The fits are overlaid on the histograms in Fig.~\ref{fig:histograms}.

We see that the exponential fits describe the data well over the fitted range, though are visibly superior for vortex matching chains. Indeed, aside from a chain length of one, the VM fits accurately capture the entire data set, whereas the VC fits suffer from a deficiency at short lengths that is especially apparent on the finer lattices.

\begin{table}
	\centering
	\caption{\label{tab:probabilities} The dimensionless probabilities ($q$) and physical rates per unit length ($\lambda$) for VC and VM chains to split into two elementary vortices at a point along the chain. The probabilities and rates are marginally larger for VM chains, constituting a contributing factor to the smaller VM chain lengths, though this distinction is only statistically significant on our finest lattice. Similar to the average chain lengths, the physical splitting rates have a mild lattice-spacing dependence that admits a linear continuum extrapolation with $a^2$. The resulting $a = 0$ values still possess the larger rate for VM objects, solidifying this as a physical discrepancy.}
	\begin{ruledtabular}
		\begin{tabular}{c|S[table-format=1.7]S[table-format=1.7]|S[table-format=1.6]S[table-format=1.6]}
			$a \, (\sigma^{-1/2})$ & \multicolumn{1}{c}{$q_\mathrm{VC}$} & \multicolumn{1}{c|}{$q_\mathrm{VM}$} & \multicolumn{1}{c}{$\lambda_\mathrm{VC} \, (\sigma^{1/2})$} & \multicolumn{1}{c}{$\lambda_\mathrm{VM} \, (\sigma^{1/2})$} \\
			\colrule
			$0.239$ & 0.2631(47) & 0.2654(48) & 1.102(20) & 1.112(20) \\
			$0.149$ & 0.1832(19) & 0.1854(17) & 1.228(13) & 1.243(11) \\
			$0.103$ & 0.1295(8)  & 0.1315(8)  & 1.259(7)  & 1.278(8) \\
			\colrule
			$0$ & \multicolumn{1}{c}{--} & \multicolumn{1}{c|}{--} & 1.296(10) & 1.317(10)
		\end{tabular}
	\end{ruledtabular}
\end{table}
The calculated probabilities are provided in Table~\ref{tab:probabilities}. As with all previous quantities, these are obtained through implementing the exponential fits on 100 bootstrap ensembles. The resulting uncertainties are particularly large on the coarser ensembles and lessen as the lattice is made finer, with the larger lattice volumes inducing improved statistics for the fits. We find a higher splitting probability for VM chains compared to VC, though the two values are only distinguishable within statistical uncertainty on our finest lattice. Nevertheless, it seems likely this is a genuine difference that factors into the shorter average VM chain length.

\begin{figure}
	\centering
	\includegraphics[width=\linewidth]{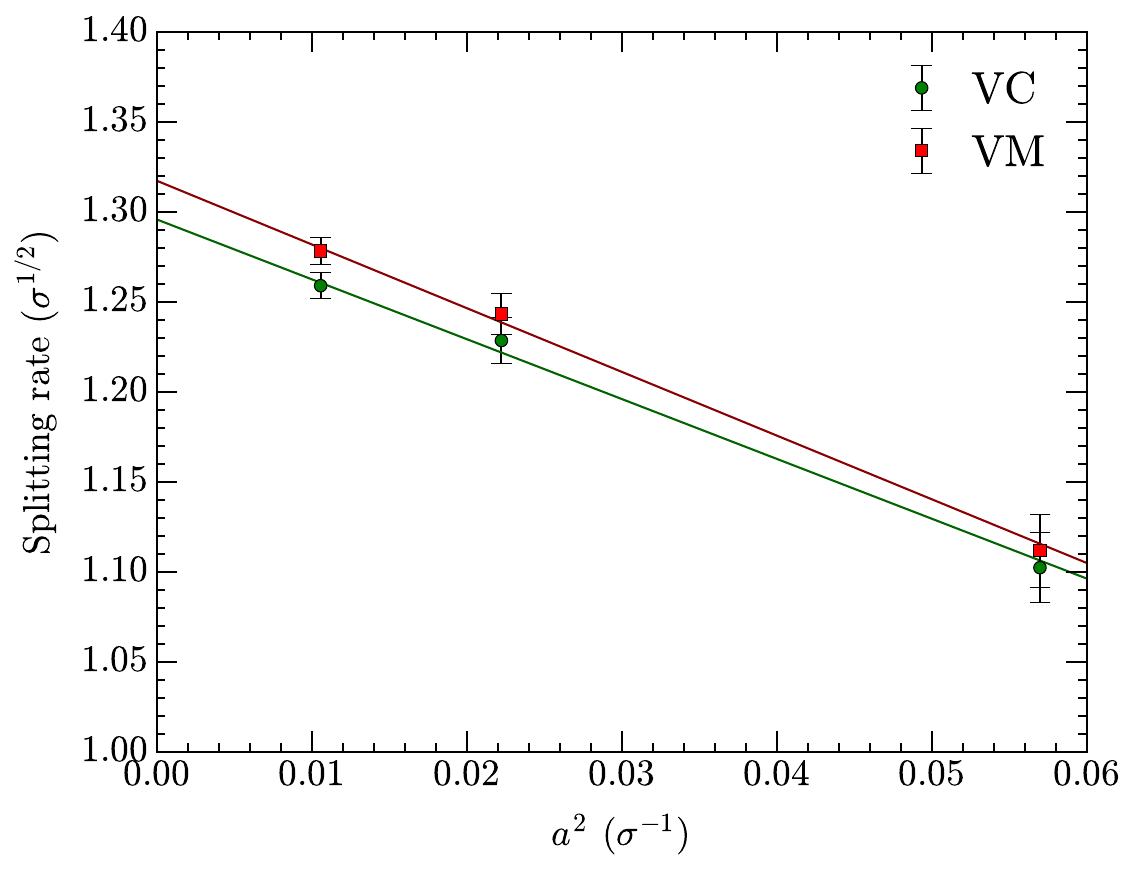}
	
	\vspace{-0.5em}
	
	\caption{\label{fig:splitting_rate} The physical splitting rates for VC and VM objects as a function of $a^2$, with linear extrapolations to the continuum limit. The fits are seen to capture the data trend, with reduced $\chi^2$ values of $\chi^2/\mathrm{d.o.f.} = 0.368$ for VC and $0.252$ for VM. These are affected by the larger uncertainties. The greater splitting probability for VM chains elucidated by our finest lattice is seen to persist in the continuum limit (at $a = 0$).}
\end{figure}

Moreover, the probabilities decrease rapidly with the lattice spacing. This is unsurprising given that it is the physical chain lengths that are approximately scale invariant. As the average chain length in lattice units diverges, the dimensionless splitting probability necessarily decays to zero in turn. Consequently, a more interesting comparison across different lattice spacings is obtained by computing a physical splitting \textit{rate} (i.e.\ probability per unit length) as $\lambda = q/a$. Looking at these values in Table~\ref{tab:probabilities}, we see a gentle increase as the lattice spacing is reduced. This is the parallel observation to the slight decrease in average chain length with the lattice spacing, and again motivates performing linear extrapolations in $a^2$ to the continuum limit. These are shown in Fig.~\ref{fig:splitting_rate}.

We see that the simple linear fits accurately describe the data, with $\chi^2/\mathrm{d.o.f.}$ values in this case of $0.368$ for VC and $0.252$ for VM. These are considerably smaller than for the average chain lengths, which can be primarily attributed to the larger statistical uncertainties, especially on the coarsest lattice. As before, we read off the values at $a = 0$; these are supplied in Table~\ref{tab:probabilities}.

Crucially, the statistically significant discrepancy between VC and VM splitting rates does survive the continuum limit, suggesting this is a physical difference. Namely, there is a greater probability for VM chains to split as they propagate through spacetime over VC. As described, this is inherently connected to the shorter average VM chain length. Combined, these findings illustrate the fascinating differences between the various aspects of center-vortex geometry in $\mathrm{SU}(4)$.

\subsection{Comparison to the SU(4) random vortex world-surface model} \label{subsec:comparison}
The numerical study presented herein lays the groundwork for effective descriptions of the $\mathrm{SU}(N)$ Yang-Mills vacuum in terms of both oriented and nonoriented center vortices. Understanding the interplay between elementary and doubly charged vortices, along with their various properties, is crucial for establishing them as candidates to describe confinement. In previous works, the $\mathrm{SU}(4)$ phenomenology was modeled in terms of random surfaces, which provided valuable insight into the large-scale behavior of center vortices \cite{Engelhardt:2005qu}. To this end, a lattice with a physical finite spacing of the order of the transverse size of thick center vortices was employed. The aim was to accommodate the string tensions $\sigma_1$ and $\sigma_2$, for quark representations with $N$-ality $k=1$ and $k=2$, respectively, along with the first-order phase transition observed in $\mathrm{SU}(4)$ YM theory.

In a model only based on curvature and area terms, increasing $\sigma_2/\sigma_1$ beyond $1.17$ in an effort to approach the observed Yang-Mills value of $\sigma_2/\sigma_1 \simeq 1.375$ \cite{Lucini:2004my, Athenodorou:2021qvs} led to the loss of the first-order transition. However, introducing two independent curvature parameters $c_1$ and $c_2$ for singly and doubly charged vortices, as well as an explicit branching term, the desired properties were successfully reproduced. The need for these components is consistent with the vortex network revealed by our simulations extrapolated to the continuum.  Moreover, if the correlation between curvature and area exhibited by the $\mathrm{SU}(2)$ and $\mathrm{SU}(3)$ random world-surface models \cite{Engelhardt:2003wm} persists for $\mathrm{SU}(4)$, the larger adjusted $c_2$ compared to $c_1$ might be related to  the shorter doubly charged chains, relative to the elementary ones, we observed at fixed-time slices.

On the other hand, as pointed out in Ref.~\cite{Engelhardt:2005qu}, the need for explicit branching and two independent curvature parameters (no area terms required) reduces the model’s predictive power. As noted,  this would be related to flux directions becoming increasingly relevant in color space with growing $N$. This contrasts with the situation in $\mathrm{SU}(3)$ \cite{Engelhardt:2003wm}, where such a term is not necessary to reproduce the first-order phase transition. In that case, branching is governed indirectly by the vortex dynamics and the entropy of branched random surfaces. 

\section{Conclusion} \label{sec:conclusion}
In this work, we have studied the geometry of center vortices in the ground-state fields of $\mathrm{SU}(4)$ Yang-Mills theory, with a focus on the newly identified physically distinct doubly charged vortices compared to $\mathrm{SU}(3)$. We have utilized three different lattice spacings down to $a \simeq 0.044\,\text{fm}$ as a means to investigate the scaling of various vortex quantities and their properties in taking the continuum limit.

Initially, visualizations of the center-vortex structure revealed that vortex matter is dominated by elementary center vortices and also displayed the different types of doubly charged objects that comprise the network. These include ``vortex convergences" (VC), which look like two elementary vortex lines that have temporarily merged together to form a doubly charged chain, and ``vortex matchings" (VM) that see both ends of the chain formed by convergent (or emergent) elementary vortices. A close inspection of the visualizations revealed that VM chains can be interpreted as extended center monopoles, in the sense that they can be thought of as monopoles where the four emanating lines coalesce in pairs over some extent. The third possibility is a secondary cluster that consists entirely of doubly charged vortex loops (L2).

An algorithm was developed to classify every doubly charged chain in three-dimensional slices of the lattice as one of these three types, paving the way to a comprehensive quantitative analysis. Investigating the relative proportions of each type of charged object showed that VC and VM chains are not created equally, with VM objects slightly favored on all three lattice spacings. This was attributed to the persistence of short VM chains across multiple three-dimensional slices. Due to center charge conservation, a VM chain can shorten down to the limit of a center monopole, which may nucleate a VM chain in a subsequent slice but cannot split into two separate parts. This is in contrast to short VC objects that can easily dissolve into their constituent elementary vortex lines, conserving charge. In addition, L2 objects were found to be rare and primarily encompassed simple $1 \times 1$ doubly charged vortex loops.

Thereafter, a detailed study into the lengths of doubly charged chains was conducted, beginning with their average lengths. This revealed that VM chains tend to be shorter than VC, possibly due to the persistence of small VM objects affecting their average lengths in the bulk. Furthermore, a soft lattice-spacing dependence of their physical lengths was observed, motivating an extrapolation to the continuum limit. Linear fits in $a^2$ were performed and found to adequately describe the data within statistical uncertainty. Extracting the values at $a = 0$ corroborated that the discrepancy between VC and VM chain lengths is physical, persisting in the continuum.

Finally, the intrinsic distributions of chain lengths were examined and observed to be strongly exponential at large chain lengths for both VC and VM objects. This is barring a preference for VM chains to have a length of one lattice unit regardless of the spacing, certainly a contributing factor to the shorter average VM chain length. The exponential distribution admits the definition of a constant ``splitting probability" for a doubly charged chain to split into two elementary vortices at a given point along the chain, extracted by performing exponential fits to the distributions. This was found to be larger for VM chains and verified to survive the continuum limit, once again congruent with their shorter average length. The differences are milder than those observed in the relative number of VC and VM chains, and in their average lengths, reaching statistical significance only on our finest lattice.

In the future, it would be beneficial to map out the \textit{areas} formed by doubly charged vortices in the full four dimensions, as opposed to lines in three-dimensional slices. It will be interesting to understand the detailed nature of the various aspects of the vortex sheet in $\mathrm{SU}(4)$. The algorithm needed for such an analysis would no doubt be more sophisticated to account for the complex behavior of the vortex sheet in four dimensions.

In a future contribution, we will revisit the center-vortex landscape in the context of the models introduced in Refs.~\cite{Oxman:2018dzp, Junior:2022bol, Junior:2024urr} for oriented and nonoriented center-vortex fluxes, which accommodate $N$-ality together with the formation of a confining flux tube. In particular, a minimal modeling based on Abelian-projected collimated fluxes was presented in Ref.~\cite{Junior:2022bol} (wave functional) and Ref.~\cite{Junior:2024urr} (Weingarten representation). The simplest version accommodates Casimir scaling, yielding $\sigma_2/\sigma_1 = 1.333$, which is already close to the measured value in $\mathrm{SU}(4)$ YM theory. These models are built solely from elementary Cartan fluxes, so it would be particularly interesting to observe doubly charged chains as emergent phenomena, with their properties derived from those of elementary center vortices. These investigations are currently underway.

\begin{acknowledgments}
This work was supported with supercomputing resources provided by the Phoenix High Performance Computing (HPC) service at the University of Adelaide. This research was undertaken with the assistance of resources and services from the National Computational Infrastructure (NCI), which is supported by the Australian Government. This research was supported by the Australian Research Council through Grant No.\ DP210103706. Financial support from the Brazilian agency CNPq under Contract No.\ 309971/2021-7 is also gratefully acknowledged.
\end{acknowledgments}

\bibliography{main}

\providecommand{\noopsort}[1]{}\providecommand{\singleletter}[1]{#1}%
\begin{thebibliography}{55}%
\makeatletter
\providecommand \@ifxundefined [1]{%
 \@ifx{#1\undefined}
}%
\providecommand \@ifnum [1]{%
 \ifnum #1\expandafter \@firstoftwo
 \else \expandafter \@secondoftwo
 \fi
}%
\providecommand \@ifx [1]{%
 \ifx #1\expandafter \@firstoftwo
 \else \expandafter \@secondoftwo
 \fi
}%
\providecommand \natexlab [1]{#1}%
\providecommand \enquote  [1]{``#1''}%
\providecommand \bibnamefont  [1]{#1}%
\providecommand \bibfnamefont [1]{#1}%
\providecommand \citenamefont [1]{#1}%
\providecommand \href@noop [0]{\@secondoftwo}%
\providecommand \href [0]{\begingroup \@sanitize@url \@href}%
\providecommand \@href[1]{\@@startlink{#1}\@@href}%
\providecommand \@@href[1]{\endgroup#1\@@endlink}%
\providecommand \@sanitize@url [0]{\catcode `\\12\catcode `\$12\catcode
  `\&12\catcode `\#12\catcode `\^12\catcode `\_12\catcode `\%12\relax}%
\providecommand \@@startlink[1]{}%
\providecommand \@@endlink[0]{}%
\providecommand \url  [0]{\begingroup\@sanitize@url \@url }%
\providecommand \@url [1]{\endgroup\@href {#1}{\urlprefix }}%
\providecommand \urlprefix  [0]{URL }%
\providecommand \Eprint [0]{\href }%
\providecommand \doibase [0]{https://doi.org/}%
\providecommand \selectlanguage [0]{\@gobble}%
\providecommand \bibinfo  [0]{\@secondoftwo}%
\providecommand \bibfield  [0]{\@secondoftwo}%
\providecommand \translation [1]{[#1]}%
\providecommand \BibitemOpen [0]{}%
\providecommand \bibitemStop [0]{}%
\providecommand \bibitemNoStop [0]{.\EOS\space}%
\providecommand \EOS [0]{\spacefactor3000\relax}%
\providecommand \BibitemShut  [1]{\csname bibitem#1\endcsname}%
\let\auto@bib@innerbib\@empty
\bibitem [{\citenamefont {Mandelstam}(1976)}]{Mandelstam:1974pi}%
  \BibitemOpen
  \bibfield  {author} {\bibinfo {author} {\bibfnamefont {S.}~\bibnamefont
  {Mandelstam}},\ }\bibfield  {title} {\bibinfo {title} {{II. Vortices and
  quark confinement in non-Abelian gauge theories}},\ }\href
  {https://doi.org/10.1016/0370-1573(76)90043-0} {\bibfield  {journal}
  {\bibinfo  {journal} {Phys. Rept.}\ }\textbf {\bibinfo {volume} {23}},\
  \bibinfo {pages} {245} (\bibinfo {year} {1976})}\BibitemShut {NoStop}%
\bibitem [{\citenamefont {'t~Hooft}(1978)}]{tHooft:1977nqb}%
  \BibitemOpen
  \bibfield  {author} {\bibinfo {author} {\bibfnamefont {G.}~\bibnamefont
  {'t~Hooft}},\ }\bibfield  {title} {\bibinfo {title} {{On the phase transition
  towards permanent quark confinement}},\ }\href
  {https://doi.org/10.1016/0550-3213(78)90153-0} {\bibfield  {journal}
  {\bibinfo  {journal} {Nucl. Phys. B}\ }\textbf {\bibinfo {volume} {138}},\
  \bibinfo {pages} {1} (\bibinfo {year} {1978})}\BibitemShut {NoStop}%
\bibitem [{\citenamefont {'t~Hooft}(1979)}]{tHooft:1979rtg}%
  \BibitemOpen
  \bibfield  {author} {\bibinfo {author} {\bibfnamefont {G.}~\bibnamefont
  {'t~Hooft}},\ }\bibfield  {title} {\bibinfo {title} {{A property of electric
  and magnetic flux in non-Abelian gauge theories}},\ }\href
  {https://doi.org/10.1016/0550-3213(79)90595-9} {\bibfield  {journal}
  {\bibinfo  {journal} {Nucl. Phys. B}\ }\textbf {\bibinfo {volume} {153}},\
  \bibinfo {pages} {141} (\bibinfo {year} {1979})}\BibitemShut {NoStop}%
\bibitem [{\citenamefont {Mack}\ and\ \citenamefont
  {Petkova}(1979)}]{Mack:1978rq}%
  \BibitemOpen
  \bibfield  {author} {\bibinfo {author} {\bibfnamefont {G.}~\bibnamefont
  {Mack}}\ and\ \bibinfo {author} {\bibfnamefont {V.~B.}\ \bibnamefont
  {Petkova}},\ }\bibfield  {title} {\bibinfo {title} {{Comparison of lattice
  gauge theories with gauge groups $Z_2$ and $SU(2)$}},\ }\href
  {https://doi.org/10.1016/0003-4916(79)90346-4} {\bibfield  {journal}
  {\bibinfo  {journal} {Annals Phys.}\ }\textbf {\bibinfo {volume} {123}},\
  \bibinfo {pages} {442} (\bibinfo {year} {1979})}\BibitemShut {NoStop}%
\bibitem [{\citenamefont {Nielsen}\ and\ \citenamefont
  {Olesen}(1979)}]{Nielsen:1979xu}%
  \BibitemOpen
  \bibfield  {author} {\bibinfo {author} {\bibfnamefont {H.~B.}\ \bibnamefont
  {Nielsen}}\ and\ \bibinfo {author} {\bibfnamefont {P.}~\bibnamefont
  {Olesen}},\ }\bibfield  {title} {\bibinfo {title} {{A quantum liquid model
  for the QCD vacuum: Gauge and rotational invariance of domained and quantized
  homogeneous color fields}},\ }\href
  {https://doi.org/10.1016/0550-3213(79)90065-8} {\bibfield  {journal}
  {\bibinfo  {journal} {Nucl. Phys. B}\ }\textbf {\bibinfo {volume} {160}},\
  \bibinfo {pages} {380} (\bibinfo {year} {1979})}\BibitemShut {NoStop}%
\bibitem [{\citenamefont {Del~Debbio}\ \emph
  {et~al.}(1997{\natexlab{a}})\citenamefont {Del~Debbio}, \citenamefont
  {Faber}, \citenamefont {Greensite},\ and\ \citenamefont
  {Olejn{\'i}k}}]{DelDebbio:1996lih}%
  \BibitemOpen
  \bibfield  {author} {\bibinfo {author} {\bibfnamefont {L.}~\bibnamefont
  {Del~Debbio}}, \bibinfo {author} {\bibfnamefont {M.}~\bibnamefont {Faber}},
  \bibinfo {author} {\bibfnamefont {J.}~\bibnamefont {Greensite}},\ and\
  \bibinfo {author} {\bibfnamefont {{\v S}.}~\bibnamefont {Olejn{\'i}k}},\
  }\bibfield  {title} {\bibinfo {title} {{Center dominance and $Z_2$ vortices
  in SU(2) lattice gauge theory}},\ }\href
  {https://doi.org/10.1103/PhysRevD.55.2298} {\bibfield  {journal} {\bibinfo
  {journal} {Phys. Rev. D}\ }\textbf {\bibinfo {volume} {55}},\ \bibinfo
  {pages} {2298} (\bibinfo {year} {1997}{\natexlab{a}})},\ \Eprint
  {https://arxiv.org/abs/hep-lat/9610005} {arXiv:hep-lat/9610005} \BibitemShut
  {NoStop}%
\bibitem [{\citenamefont {Del~Debbio}\ \emph
  {et~al.}(1998{\natexlab{a}})\citenamefont {Del~Debbio}, \citenamefont
  {Faber}, \citenamefont {Greensite},\ and\ \citenamefont
  {Olejn{\'i}k}}]{DelDebbio:1997ep}%
  \BibitemOpen
  \bibfield  {author} {\bibinfo {author} {\bibfnamefont {L.}~\bibnamefont
  {Del~Debbio}}, \bibinfo {author} {\bibfnamefont {M.}~\bibnamefont {Faber}},
  \bibinfo {author} {\bibfnamefont {J.}~\bibnamefont {Greensite}},\ and\
  \bibinfo {author} {\bibfnamefont {{\v S}.}~\bibnamefont {Olejn{\'i}k}},\
  }\bibfield  {title} {\bibinfo {title} {{Center vortices and the asymptotic
  string tension}},\ }\href {https://doi.org/10.1016/S0920-5632(97)00831-1}
  {\bibfield  {journal} {\bibinfo  {journal} {Nucl. Phys. B Proc. Suppl.}\
  }\textbf {\bibinfo {volume} {63}},\ \bibinfo {pages} {552} (\bibinfo {year}
  {1998}{\natexlab{a}})},\ \Eprint {https://arxiv.org/abs/hep-lat/9709032}
  {arXiv:hep-lat/9709032} \BibitemShut {NoStop}%
\bibitem [{\citenamefont {Langfeld}\ \emph {et~al.}(1998)\citenamefont
  {Langfeld}, \citenamefont {Reinhardt},\ and\ \citenamefont
  {Tennert}}]{Langfeld:1997jx}%
  \BibitemOpen
  \bibfield  {author} {\bibinfo {author} {\bibfnamefont {K.}~\bibnamefont
  {Langfeld}}, \bibinfo {author} {\bibfnamefont {H.}~\bibnamefont
  {Reinhardt}},\ and\ \bibinfo {author} {\bibfnamefont {O.}~\bibnamefont
  {Tennert}},\ }\bibfield  {title} {\bibinfo {title} {{Confinement and scaling
  of the vortex vacuum of SU(2) lattice gauge theory}},\ }\href
  {https://doi.org/10.1016/S0370-2693(97)01435-4} {\bibfield  {journal}
  {\bibinfo  {journal} {Phys. Lett. B}\ }\textbf {\bibinfo {volume} {419}},\
  \bibinfo {pages} {317} (\bibinfo {year} {1998})},\ \Eprint
  {https://arxiv.org/abs/hep-lat/9710068} {arXiv:hep-lat/9710068} \BibitemShut
  {NoStop}%
\bibitem [{\citenamefont {Del~Debbio}\ \emph
  {et~al.}(1998{\natexlab{b}})\citenamefont {Del~Debbio}, \citenamefont
  {Faber}, \citenamefont {Giedt}, \citenamefont {Greensite},\ and\
  \citenamefont {Olejn{\'i}k}}]{DelDebbio:1998luz}%
  \BibitemOpen
  \bibfield  {author} {\bibinfo {author} {\bibfnamefont {L.}~\bibnamefont
  {Del~Debbio}}, \bibinfo {author} {\bibfnamefont {M.}~\bibnamefont {Faber}},
  \bibinfo {author} {\bibfnamefont {J.}~\bibnamefont {Giedt}}, \bibinfo
  {author} {\bibfnamefont {J.}~\bibnamefont {Greensite}},\ and\ \bibinfo
  {author} {\bibfnamefont {{\v S}.}~\bibnamefont {Olejn{\'i}k}},\ }\bibfield
  {title} {\bibinfo {title} {{Detection of center vortices in the lattice
  Yang-Mills vacuum}},\ }\href {https://doi.org/10.1103/PhysRevD.58.094501}
  {\bibfield  {journal} {\bibinfo  {journal} {Phys. Rev. D}\ }\textbf {\bibinfo
  {volume} {58}},\ \bibinfo {pages} {094501} (\bibinfo {year}
  {1998}{\natexlab{b}})},\ \Eprint {https://arxiv.org/abs/hep-lat/9801027}
  {arXiv:hep-lat/9801027} \BibitemShut {NoStop}%
\bibitem [{\citenamefont {Faber}\ \emph {et~al.}(1998)\citenamefont {Faber},
  \citenamefont {Greensite},\ and\ \citenamefont {Olejn{\'i}k}}]{Faber:1997rp}%
  \BibitemOpen
  \bibfield  {author} {\bibinfo {author} {\bibfnamefont {M.}~\bibnamefont
  {Faber}}, \bibinfo {author} {\bibfnamefont {J.}~\bibnamefont {Greensite}},\
  and\ \bibinfo {author} {\bibfnamefont {{\v S}.}~\bibnamefont {Olejn{\'i}k}},\
  }\bibfield  {title} {\bibinfo {title} {{Casimir scaling from center vortices:
  Towards an understanding of the adjoint string tension}},\ }\href
  {https://doi.org/10.1103/PhysRevD.57.2603} {\bibfield  {journal} {\bibinfo
  {journal} {Phys. Rev. D}\ }\textbf {\bibinfo {volume} {57}},\ \bibinfo
  {pages} {2603} (\bibinfo {year} {1998})},\ \Eprint
  {https://arxiv.org/abs/hep-lat/9710039} {arXiv:hep-lat/9710039} \BibitemShut
  {NoStop}%
\bibitem [{\citenamefont {Faber}\ \emph
  {et~al.}(1999{\natexlab{a}})\citenamefont {Faber}, \citenamefont
  {Greensite},\ and\ \citenamefont {Olejn{\'i}k}}]{Faber:1998qn}%
  \BibitemOpen
  \bibfield  {author} {\bibinfo {author} {\bibfnamefont {M.}~\bibnamefont
  {Faber}}, \bibinfo {author} {\bibfnamefont {J.}~\bibnamefont {Greensite}},\
  and\ \bibinfo {author} {\bibfnamefont {{\v S}.}~\bibnamefont {Olejn{\'i}k}},\
  }\bibfield  {title} {\bibinfo {title} {{Evidence for a center vortex origin
  of the adjoint string tension}},\ }\href@noop {} {\bibfield  {journal}
  {\bibinfo  {journal} {Acta Phys. Slov.}\ }\textbf {\bibinfo {volume} {49}},\
  \bibinfo {pages} {177} (\bibinfo {year} {1999}{\natexlab{a}})},\ \Eprint
  {https://arxiv.org/abs/hep-lat/9807008} {arXiv:hep-lat/9807008} \BibitemShut
  {NoStop}%
\bibitem [{\citenamefont {Kov{\'a}cs}\ and\ \citenamefont
  {Tomboulis}(1998)}]{Kovacs:1998xm}%
  \BibitemOpen
  \bibfield  {author} {\bibinfo {author} {\bibfnamefont {T.~G.}\ \bibnamefont
  {Kov{\'a}cs}}\ and\ \bibinfo {author} {\bibfnamefont {E.~T.}\ \bibnamefont
  {Tomboulis}},\ }\bibfield  {title} {\bibinfo {title} {{Vortices and
  confinement at weak coupling}},\ }\href
  {https://doi.org/10.1103/PhysRevD.57.4054} {\bibfield  {journal} {\bibinfo
  {journal} {Phys. Rev. D}\ }\textbf {\bibinfo {volume} {57}},\ \bibinfo
  {pages} {4054} (\bibinfo {year} {1998})},\ \Eprint
  {https://arxiv.org/abs/hep-lat/9711009} {arXiv:hep-lat/9711009} \BibitemShut
  {NoStop}%
\bibitem [{\citenamefont {Langfeld}\ \emph {et~al.}(1999)\citenamefont
  {Langfeld}, \citenamefont {Tennert}, \citenamefont {Engelhardt},\ and\
  \citenamefont {Reinhardt}}]{Langfeld:1998cz}%
  \BibitemOpen
  \bibfield  {author} {\bibinfo {author} {\bibfnamefont {K.}~\bibnamefont
  {Langfeld}}, \bibinfo {author} {\bibfnamefont {O.}~\bibnamefont {Tennert}},
  \bibinfo {author} {\bibfnamefont {M.}~\bibnamefont {Engelhardt}},\ and\
  \bibinfo {author} {\bibfnamefont {H.}~\bibnamefont {Reinhardt}},\ }\bibfield
  {title} {\bibinfo {title} {{Center vortices of Yang-Mills theory at finite
  temperatures}},\ }\href {https://doi.org/10.1016/S0370-2693(99)00252-X}
  {\bibfield  {journal} {\bibinfo  {journal} {Phys. Lett. B}\ }\textbf
  {\bibinfo {volume} {452}},\ \bibinfo {pages} {301} (\bibinfo {year}
  {1999})},\ \Eprint {https://arxiv.org/abs/hep-lat/9805002}
  {arXiv:hep-lat/9805002} \BibitemShut {NoStop}%
\bibitem [{\citenamefont {Bertle}\ \emph {et~al.}(1999)\citenamefont {Bertle},
  \citenamefont {Faber}, \citenamefont {Greensite},\ and\ \citenamefont
  {Olejn{\'i}k}}]{Bertle:1999tw}%
  \BibitemOpen
  \bibfield  {author} {\bibinfo {author} {\bibfnamefont {R.}~\bibnamefont
  {Bertle}}, \bibinfo {author} {\bibfnamefont {M.}~\bibnamefont {Faber}},
  \bibinfo {author} {\bibfnamefont {J.}~\bibnamefont {Greensite}},\ and\
  \bibinfo {author} {\bibfnamefont {{\v S}.}~\bibnamefont {Olejn{\'i}k}},\
  }\bibfield  {title} {\bibinfo {title} {{The structure of projected center
  vortices in lattice gauge theory}},\ }\href
  {https://doi.org/10.1088/1126-6708/1999/03/019} {\bibfield  {journal}
  {\bibinfo  {journal} {J. High Energy Phys.}\ }\textbf {\bibinfo {volume}
  {03}},\ \bibinfo {pages} {019}},\ \Eprint
  {https://arxiv.org/abs/hep-lat/9903023} {arXiv:hep-lat/9903023} \BibitemShut
  {NoStop}%
\bibitem [{\citenamefont {Engelhardt}\ \emph {et~al.}(2000)\citenamefont
  {Engelhardt}, \citenamefont {Langfeld}, \citenamefont {Reinhardt},\ and\
  \citenamefont {Tennert}}]{Engelhardt:1999fd}%
  \BibitemOpen
  \bibfield  {author} {\bibinfo {author} {\bibfnamefont {M.}~\bibnamefont
  {Engelhardt}}, \bibinfo {author} {\bibfnamefont {K.}~\bibnamefont
  {Langfeld}}, \bibinfo {author} {\bibfnamefont {H.}~\bibnamefont
  {Reinhardt}},\ and\ \bibinfo {author} {\bibfnamefont {O.}~\bibnamefont
  {Tennert}},\ }\bibfield  {title} {\bibinfo {title} {{Deconfinement in SU(2)
  Yang-Mills theory as a center vortex percolation transition}},\ }\href
  {https://doi.org/10.1103/PhysRevD.61.054504} {\bibfield  {journal} {\bibinfo
  {journal} {Phys. Rev. D}\ }\textbf {\bibinfo {volume} {61}},\ \bibinfo
  {pages} {054504} (\bibinfo {year} {2000})},\ \Eprint
  {https://arxiv.org/abs/hep-lat/9904004} {arXiv:hep-lat/9904004} \BibitemShut
  {NoStop}%
\bibitem [{\citenamefont {Engelhardt}\ and\ \citenamefont
  {Reinhardt}(2000)}]{Engelhardt:1999wr}%
  \BibitemOpen
  \bibfield  {author} {\bibinfo {author} {\bibfnamefont {M.}~\bibnamefont
  {Engelhardt}}\ and\ \bibinfo {author} {\bibfnamefont {H.}~\bibnamefont
  {Reinhardt}},\ }\bibfield  {title} {\bibinfo {title} {{Center vortex model
  for the infrared sector of Yang-Mills theory --- confinement and
  deconfinement}},\ }\href {https://doi.org/10.1016/S0550-3213(00)00445-4}
  {\bibfield  {journal} {\bibinfo  {journal} {Nucl. Phys. B}\ }\textbf
  {\bibinfo {volume} {585}},\ \bibinfo {pages} {591} (\bibinfo {year}
  {2000})},\ \Eprint {https://arxiv.org/abs/hep-lat/9912003}
  {arXiv:hep-lat/9912003} \BibitemShut {NoStop}%
\bibitem [{\citenamefont {Faber}\ \emph {et~al.}(2000)\citenamefont {Faber},
  \citenamefont {Greensite},\ and\ \citenamefont {Olejn{\'i}k}}]{Faber:1999sq}%
  \BibitemOpen
  \bibfield  {author} {\bibinfo {author} {\bibfnamefont {M.}~\bibnamefont
  {Faber}}, \bibinfo {author} {\bibfnamefont {J.}~\bibnamefont {Greensite}},\
  and\ \bibinfo {author} {\bibfnamefont {{\v S}.}~\bibnamefont {Olejn{\'i}k}},\
  }\bibfield  {title} {\bibinfo {title} {{First evidence for center dominance
  in SU(3) lattice gauge theory}},\ }\href
  {https://doi.org/10.1016/S0370-2693(00)00013-7} {\bibfield  {journal}
  {\bibinfo  {journal} {Phys. Lett. B}\ }\textbf {\bibinfo {volume} {474}},\
  \bibinfo {pages} {177} (\bibinfo {year} {2000})},\ \Eprint
  {https://arxiv.org/abs/hep-lat/9911006} {arXiv:hep-lat/9911006} \BibitemShut
  {NoStop}%
\bibitem [{\citenamefont {de~Forcrand}\ and\ \citenamefont
  {D'Elia}(1999)}]{deForcrand:1999our}%
  \BibitemOpen
  \bibfield  {author} {\bibinfo {author} {\bibfnamefont {P.}~\bibnamefont
  {de~Forcrand}}\ and\ \bibinfo {author} {\bibfnamefont {M.}~\bibnamefont
  {D'Elia}},\ }\bibfield  {title} {\bibinfo {title} {{Relevance of Center
  Vortices to QCD}},\ }\href {https://doi.org/10.1103/PhysRevLett.82.4582}
  {\bibfield  {journal} {\bibinfo  {journal} {Phys. Rev. Lett.}\ }\textbf
  {\bibinfo {volume} {82}},\ \bibinfo {pages} {4582} (\bibinfo {year}
  {1999})},\ \Eprint {https://arxiv.org/abs/hep-lat/9901020}
  {arXiv:hep-lat/9901020} \BibitemShut {NoStop}%
\bibitem [{\citenamefont {Kov{\'a}cs}\ and\ \citenamefont
  {Tomboulis}(2000)}]{Kovacs:2000sy}%
  \BibitemOpen
  \bibfield  {author} {\bibinfo {author} {\bibfnamefont {T.~G.}\ \bibnamefont
  {Kov{\'a}cs}}\ and\ \bibinfo {author} {\bibfnamefont {E.~T.}\ \bibnamefont
  {Tomboulis}},\ }\bibfield  {title} {\bibinfo {title} {{Computation of the
  Vortex Free Energy in SU(2) Gauge Theory}},\ }\href
  {https://doi.org/10.1103/PhysRevLett.85.704} {\bibfield  {journal} {\bibinfo
  {journal} {Phys. Rev. Lett.}\ }\textbf {\bibinfo {volume} {85}},\ \bibinfo
  {pages} {704} (\bibinfo {year} {2000})},\ \Eprint
  {https://arxiv.org/abs/hep-lat/0002004} {arXiv:hep-lat/0002004} \BibitemShut
  {NoStop}%
\bibitem [{\citenamefont {Langfeld}\ \emph {et~al.}(2002)\citenamefont
  {Langfeld}, \citenamefont {Reinhardt},\ and\ \citenamefont
  {Gattnar}}]{Langfeld:2001cz}%
  \BibitemOpen
  \bibfield  {author} {\bibinfo {author} {\bibfnamefont {K.}~\bibnamefont
  {Langfeld}}, \bibinfo {author} {\bibfnamefont {H.}~\bibnamefont
  {Reinhardt}},\ and\ \bibinfo {author} {\bibfnamefont {J.}~\bibnamefont
  {Gattnar}},\ }\bibfield  {title} {\bibinfo {title} {{Gluon propagator and
  quark confinement}},\ }\href {https://doi.org/10.1016/S0550-3213(01)00574-0}
  {\bibfield  {journal} {\bibinfo  {journal} {Nucl. Phys. B}\ }\textbf
  {\bibinfo {volume} {621}},\ \bibinfo {pages} {131} (\bibinfo {year}
  {2002})},\ \Eprint {https://arxiv.org/abs/hep-ph/0107141}
  {arXiv:hep-ph/0107141} \BibitemShut {NoStop}%
\bibitem [{\citenamefont {Langfeld}(2004)}]{Langfeld:2003ev}%
  \BibitemOpen
  \bibfield  {author} {\bibinfo {author} {\bibfnamefont {K.}~\bibnamefont
  {Langfeld}},\ }\bibfield  {title} {\bibinfo {title} {{Vortex structures in
  pure SU(3) lattice gauge theory}},\ }\href
  {https://doi.org/10.1103/PhysRevD.69.014503} {\bibfield  {journal} {\bibinfo
  {journal} {Phys. Rev. D}\ }\textbf {\bibinfo {volume} {69}},\ \bibinfo
  {pages} {014503} (\bibinfo {year} {2004})},\ \Eprint
  {https://arxiv.org/abs/hep-lat/0307030} {arXiv:hep-lat/0307030} \BibitemShut
  {NoStop}%
\bibitem [{\citenamefont {Greensite}(2003)}]{Greensite:2003bk}%
  \BibitemOpen
  \bibfield  {author} {\bibinfo {author} {\bibfnamefont {J.}~\bibnamefont
  {Greensite}},\ }\bibfield  {title} {\bibinfo {title} {{The confinement
  problem in lattice gauge theory}},\ }\href
  {https://doi.org/10.1016/S0146-6410(03)90012-3} {\bibfield  {journal}
  {\bibinfo  {journal} {Prog. Part. Nucl. Phys.}\ }\textbf {\bibinfo {volume}
  {51}},\ \bibinfo {pages} {1} (\bibinfo {year} {2003})},\ \Eprint
  {https://arxiv.org/abs/hep-lat/0301023} {arXiv:hep-lat/0301023} \BibitemShut
  {NoStop}%
\bibitem [{\citenamefont {Engelhardt}\ \emph {et~al.}(2004)\citenamefont
  {Engelhardt}, \citenamefont {Quandt},\ and\ \citenamefont
  {Reinhardt}}]{Engelhardt:2003wm}%
  \BibitemOpen
  \bibfield  {author} {\bibinfo {author} {\bibfnamefont {M.}~\bibnamefont
  {Engelhardt}}, \bibinfo {author} {\bibfnamefont {M.}~\bibnamefont {Quandt}},\
  and\ \bibinfo {author} {\bibfnamefont {H.}~\bibnamefont {Reinhardt}},\
  }\bibfield  {title} {\bibinfo {title} {{Center vortex model for the infrared
  sector of $SU(3)$ Yang-Mills theory---confinement and deconfinement}},\
  }\href {https://doi.org/10.1016/j.nuclphysb.2004.02.036} {\bibfield
  {journal} {\bibinfo  {journal} {Nucl. Phys. B}\ }\textbf {\bibinfo {volume}
  {685}},\ \bibinfo {pages} {227} (\bibinfo {year} {2004})},\ \Eprint
  {https://arxiv.org/abs/hep-lat/0311029} {arXiv:hep-lat/0311029} \BibitemShut
  {NoStop}%
\bibitem [{\citenamefont {Bowman}\ \emph {et~al.}(2008)\citenamefont {Bowman},
  \citenamefont {Langfeld}, \citenamefont {Leinweber}, \citenamefont {O'~Cais},
  \citenamefont {Sternbeck}, \citenamefont {von Smekal},\ and\ \citenamefont
  {Williams}}]{Bowman:2008qd}%
  \BibitemOpen
  \bibfield  {author} {\bibinfo {author} {\bibfnamefont {P.~O.}\ \bibnamefont
  {Bowman}}, \bibinfo {author} {\bibfnamefont {K.}~\bibnamefont {Langfeld}},
  \bibinfo {author} {\bibfnamefont {D.~B.}\ \bibnamefont {Leinweber}}, \bibinfo
  {author} {\bibfnamefont {A.}~\bibnamefont {O'~Cais}}, \bibinfo {author}
  {\bibfnamefont {A.}~\bibnamefont {Sternbeck}}, \bibinfo {author}
  {\bibfnamefont {L.}~\bibnamefont {von Smekal}},\ and\ \bibinfo {author}
  {\bibfnamefont {A.~G.}\ \bibnamefont {Williams}},\ }\bibfield  {title}
  {\bibinfo {title} {{Center vortices and the quark propagator in SU(2) gauge
  theory}},\ }\href {https://doi.org/10.1103/PhysRevD.78.054509} {\bibfield
  {journal} {\bibinfo  {journal} {Phys. Rev. D}\ }\textbf {\bibinfo {volume}
  {78}},\ \bibinfo {pages} {054509} (\bibinfo {year} {2008})},\ \Eprint
  {https://arxiv.org/abs/0806.4219} {arXiv:0806.4219 [hep-lat]} \BibitemShut
  {NoStop}%
\bibitem [{\citenamefont {Bowman}\ \emph {et~al.}(2011)\citenamefont {Bowman},
  \citenamefont {Langfeld}, \citenamefont {Leinweber}, \citenamefont
  {Sternbeck}, \citenamefont {von Smekal},\ and\ \citenamefont
  {Williams}}]{Bowman:2010zr}%
  \BibitemOpen
  \bibfield  {author} {\bibinfo {author} {\bibfnamefont {P.~O.}\ \bibnamefont
  {Bowman}}, \bibinfo {author} {\bibfnamefont {K.}~\bibnamefont {Langfeld}},
  \bibinfo {author} {\bibfnamefont {D.~B.}\ \bibnamefont {Leinweber}}, \bibinfo
  {author} {\bibfnamefont {A.}~\bibnamefont {Sternbeck}}, \bibinfo {author}
  {\bibfnamefont {L.}~\bibnamefont {von Smekal}},\ and\ \bibinfo {author}
  {\bibfnamefont {A.~G.}\ \bibnamefont {Williams}},\ }\bibfield  {title}
  {\bibinfo {title} {{Role of center vortices in chiral symmetry breaking in
  SU(3) gauge theory}},\ }\href {https://doi.org/10.1103/PhysRevD.84.034501}
  {\bibfield  {journal} {\bibinfo  {journal} {Phys. Rev. D}\ }\textbf {\bibinfo
  {volume} {84}},\ \bibinfo {pages} {034501} (\bibinfo {year} {2011})},\
  \Eprint {https://arxiv.org/abs/1010.4624} {arXiv:1010.4624 [hep-lat]}
  \BibitemShut {NoStop}%
\bibitem [{\citenamefont {O'Malley}\ \emph {et~al.}(2012)\citenamefont
  {O'Malley}, \citenamefont {Kamleh}, \citenamefont {Leinweber},\ and\
  \citenamefont {Moran}}]{OMalley:2011aa}%
  \BibitemOpen
  \bibfield  {author} {\bibinfo {author} {\bibfnamefont {E.-A.}\ \bibnamefont
  {O'Malley}}, \bibinfo {author} {\bibfnamefont {W.}~\bibnamefont {Kamleh}},
  \bibinfo {author} {\bibfnamefont {D.}~\bibnamefont {Leinweber}},\ and\
  \bibinfo {author} {\bibfnamefont {P.}~\bibnamefont {Moran}},\ }\bibfield
  {title} {\bibinfo {title} {{$SU(3)$ centre vortices underpin confinement and
  dynamical chiral symmetry breaking}},\ }\href
  {https://doi.org/10.1103/PhysRevD.86.054503} {\bibfield  {journal} {\bibinfo
  {journal} {Phys. Rev. D}\ }\textbf {\bibinfo {volume} {86}},\ \bibinfo
  {pages} {054503} (\bibinfo {year} {2012})},\ \Eprint
  {https://arxiv.org/abs/1112.2490} {arXiv:1112.2490 [hep-lat]} \BibitemShut
  {NoStop}%
\bibitem [{\citenamefont {Greensite}(2017)}]{Greensite:2016pfc}%
  \BibitemOpen
  \bibfield  {author} {\bibinfo {author} {\bibfnamefont {J.}~\bibnamefont
  {Greensite}},\ }\bibfield  {title} {\bibinfo {title} {{Confinement from
  Center Vortices: A review of old and new results}},\ }\href
  {https://doi.org/10.1051/epjconf/201713701009} {\bibfield  {journal}
  {\bibinfo  {journal} {EPJ Web Conf.}\ }\textbf {\bibinfo {volume} {137}},\
  \bibinfo {pages} {01009} (\bibinfo {year} {2017})},\ \Eprint
  {https://arxiv.org/abs/1610.06221} {arXiv:1610.06221 [hep-lat]} \BibitemShut
  {NoStop}%
\bibitem [{\citenamefont {Biddle}\ \emph
  {et~al.}(2022{\natexlab{a}})\citenamefont {Biddle}, \citenamefont {Kamleh},\
  and\ \citenamefont {Leinweber}}]{Biddle:2022zgw}%
  \BibitemOpen
  \bibfield  {author} {\bibinfo {author} {\bibfnamefont {J.~C.}\ \bibnamefont
  {Biddle}}, \bibinfo {author} {\bibfnamefont {W.}~\bibnamefont {Kamleh}},\
  and\ \bibinfo {author} {\bibfnamefont {D.~B.}\ \bibnamefont {Leinweber}},\
  }\bibfield  {title} {\bibinfo {title} {{Static quark potential from center
  vortices in the presence of dynamical fermions}},\ }\href
  {https://doi.org/10.1103/PhysRevD.106.054505} {\bibfield  {journal} {\bibinfo
   {journal} {Phys. Rev. D}\ }\textbf {\bibinfo {volume} {106}},\ \bibinfo
  {pages} {054505} (\bibinfo {year} {2022}{\natexlab{a}})},\ \Eprint
  {https://arxiv.org/abs/2206.00844} {arXiv:2206.00844 [hep-lat]} \BibitemShut
  {NoStop}%
\bibitem [{\citenamefont {Biddle}\ \emph
  {et~al.}(2022{\natexlab{b}})\citenamefont {Biddle}, \citenamefont {Kamleh},\
  and\ \citenamefont {Leinweber}}]{Biddle:2022acd}%
  \BibitemOpen
  \bibfield  {author} {\bibinfo {author} {\bibfnamefont {J.~C.}\ \bibnamefont
  {Biddle}}, \bibinfo {author} {\bibfnamefont {W.}~\bibnamefont {Kamleh}},\
  and\ \bibinfo {author} {\bibfnamefont {D.~B.}\ \bibnamefont {Leinweber}},\
  }\bibfield  {title} {\bibinfo {title} {{Impact of dynamical fermions on the
  center vortex gluon propagator}},\ }\href
  {https://doi.org/10.1103/PhysRevD.106.014506} {\bibfield  {journal} {\bibinfo
   {journal} {Phys. Rev. D}\ }\textbf {\bibinfo {volume} {106}},\ \bibinfo
  {pages} {014506} (\bibinfo {year} {2022}{\natexlab{b}})},\ \Eprint
  {https://arxiv.org/abs/2206.02320} {arXiv:2206.02320 [hep-lat]} \BibitemShut
  {NoStop}%
\bibitem [{\citenamefont {Mickley}\ \emph {et~al.}(2024)\citenamefont
  {Mickley}, \citenamefont {Kamleh},\ and\ \citenamefont
  {Leinweber}}]{Mickley:2024zyg}%
  \BibitemOpen
  \bibfield  {author} {\bibinfo {author} {\bibfnamefont {J.~A.}\ \bibnamefont
  {Mickley}}, \bibinfo {author} {\bibfnamefont {W.}~\bibnamefont {Kamleh}},\
  and\ \bibinfo {author} {\bibfnamefont {D.~B.}\ \bibnamefont {Leinweber}},\
  }\bibfield  {title} {\bibinfo {title} {{Center vortex geometry at finite
  temperature}},\ }\href {https://doi.org/10.1103/PhysRevD.110.034516}
  {\bibfield  {journal} {\bibinfo  {journal} {Phys. Rev. D}\ }\textbf {\bibinfo
  {volume} {110}},\ \bibinfo {pages} {034516} (\bibinfo {year} {2024})},\
  \Eprint {https://arxiv.org/abs/2405.10670} {arXiv:2405.10670 [hep-lat]}
  \BibitemShut {NoStop}%
\bibitem [{\citenamefont {Mickley}\ \emph {et~al.}(2025)\citenamefont
  {Mickley}, \citenamefont {Allton}, \citenamefont {Bignell},\ and\
  \citenamefont {Leinweber}}]{Mickley:2024vkm}%
  \BibitemOpen
  \bibfield  {author} {\bibinfo {author} {\bibfnamefont {J.~A.}\ \bibnamefont
  {Mickley}}, \bibinfo {author} {\bibfnamefont {C.}~\bibnamefont {Allton}},
  \bibinfo {author} {\bibfnamefont {R.}~\bibnamefont {Bignell}},\ and\ \bibinfo
  {author} {\bibfnamefont {D.~B.}\ \bibnamefont {Leinweber}},\ }\bibfield
  {title} {\bibinfo {title} {{Center vortex evidence for a second
  finite-temperature QCD transition}},\ }\href
  {https://doi.org/10.1103/PhysRevD.111.034508} {\bibfield  {journal} {\bibinfo
   {journal} {Phys. Rev. D}\ }\textbf {\bibinfo {volume} {111}},\ \bibinfo
  {pages} {034508} (\bibinfo {year} {2025})},\ \Eprint
  {https://arxiv.org/abs/2411.19446} {arXiv:2411.19446 [hep-lat]} \BibitemShut
  {NoStop}%
\bibitem [{\citenamefont {Spengler}\ \emph {et~al.}(2018)\citenamefont
  {Spengler}, \citenamefont {Quandt},\ and\ \citenamefont
  {Reinhardt}}]{Spengler:2018dxt}%
  \BibitemOpen
  \bibfield  {author} {\bibinfo {author} {\bibfnamefont {F.}~\bibnamefont
  {Spengler}}, \bibinfo {author} {\bibfnamefont {M.}~\bibnamefont {Quandt}},\
  and\ \bibinfo {author} {\bibfnamefont {H.}~\bibnamefont {Reinhardt}},\
  }\bibfield  {title} {\bibinfo {title} {{Branching of center vortices in SU(3)
  lattice gauge theory}},\ }\href {https://doi.org/10.1103/PhysRevD.98.094508}
  {\bibfield  {journal} {\bibinfo  {journal} {Phys. Rev. D}\ }\textbf {\bibinfo
  {volume} {98}},\ \bibinfo {pages} {094508} (\bibinfo {year} {2018})},\
  \Eprint {https://arxiv.org/abs/1810.04072} {arXiv:1810.04072 [hep-th]}
  \BibitemShut {NoStop}%
\bibitem [{\citenamefont {Biddle}\ \emph {et~al.}(2020)\citenamefont {Biddle},
  \citenamefont {Kamleh},\ and\ \citenamefont {Leinweber}}]{Biddle:2019gke}%
  \BibitemOpen
  \bibfield  {author} {\bibinfo {author} {\bibfnamefont {J.~C.}\ \bibnamefont
  {Biddle}}, \bibinfo {author} {\bibfnamefont {W.}~\bibnamefont {Kamleh}},\
  and\ \bibinfo {author} {\bibfnamefont {D.~B.}\ \bibnamefont {Leinweber}},\
  }\bibfield  {title} {\bibinfo {title} {{Visualization of center vortex
  structure}},\ }\href {https://doi.org/10.1103/PhysRevD.102.034504} {\bibfield
   {journal} {\bibinfo  {journal} {Phys. Rev. D}\ }\textbf {\bibinfo {volume}
  {102}},\ \bibinfo {pages} {034504} (\bibinfo {year} {2020})},\ \Eprint
  {https://arxiv.org/abs/1912.09531} {arXiv:1912.09531 [hep-lat]} \BibitemShut
  {NoStop}%
\bibitem [{\citenamefont {Biddle}\ \emph {et~al.}(2023)\citenamefont {Biddle},
  \citenamefont {Kamleh},\ and\ \citenamefont {Leinweber}}]{Biddle:2023lod}%
  \BibitemOpen
  \bibfield  {author} {\bibinfo {author} {\bibfnamefont {J.~C.}\ \bibnamefont
  {Biddle}}, \bibinfo {author} {\bibfnamefont {W.}~\bibnamefont {Kamleh}},\
  and\ \bibinfo {author} {\bibfnamefont {D.~B.}\ \bibnamefont {Leinweber}},\
  }\bibfield  {title} {\bibinfo {title} {{Center vortex structure in the
  presence of dynamical fermions}},\ }\href
  {https://doi.org/10.1103/PhysRevD.107.094507} {\bibfield  {journal} {\bibinfo
   {journal} {Phys. Rev. D}\ }\textbf {\bibinfo {volume} {107}},\ \bibinfo
  {pages} {094507} (\bibinfo {year} {2023})},\ \Eprint
  {https://arxiv.org/abs/2302.05897} {arXiv:2302.05897 [hep-lat]} \BibitemShut
  {NoStop}%
\bibitem [{\citenamefont {Engelhardt}(2006)}]{Engelhardt:2005qu}%
  \BibitemOpen
  \bibfield  {author} {\bibinfo {author} {\bibfnamefont {M.}~\bibnamefont
  {Engelhardt}},\ }\bibfield  {title} {\bibinfo {title} {{Center vortex model
  for the infrared sector of $SU(4)$ Yang-Mills theory: String tensions and
  deconfinement transition}},\ }\href
  {https://doi.org/10.1103/PhysRevD.73.034015} {\bibfield  {journal} {\bibinfo
  {journal} {Phys. Rev. D}\ }\textbf {\bibinfo {volume} {73}},\ \bibinfo
  {pages} {034015} (\bibinfo {year} {2006})},\ \Eprint
  {https://arxiv.org/abs/hep-lat/0512015} {arXiv:hep-lat/0512015} \BibitemShut
  {NoStop}%
\bibitem [{\citenamefont {Del~Debbio}\ \emph
  {et~al.}(1997{\natexlab{b}})\citenamefont {Del~Debbio}, \citenamefont
  {Faber}, \citenamefont {Greensite},\ and\ \citenamefont
  {Olejnik}}]{DelDebbio:1997ke}%
  \BibitemOpen
  \bibfield  {author} {\bibinfo {author} {\bibfnamefont {L.}~\bibnamefont
  {Del~Debbio}}, \bibinfo {author} {\bibfnamefont {M.}~\bibnamefont {Faber}},
  \bibinfo {author} {\bibfnamefont {J.}~\bibnamefont {Greensite}},\ and\
  \bibinfo {author} {\bibfnamefont {S.}~\bibnamefont {Olejnik}},\ }\bibfield
  {title} {\bibinfo {title} {{Center dominance, center vortices, and
  confinement}},\ }in\ \href@noop {} {\emph {\bibinfo {booktitle} {{NATO
  Advanced Research Workshop on Theoretical Physics: New Developments in
  Quantum Field Theory}}}}\ (\bibinfo {year} {1997})\ pp.\ \bibinfo {pages}
  {47--64},\ \Eprint {https://arxiv.org/abs/hep-lat/9708023}
  {arXiv:hep-lat/9708023} \BibitemShut {NoStop}%
\bibitem [{\citenamefont {Ambjorn}\ \emph {et~al.}(2000)\citenamefont
  {Ambjorn}, \citenamefont {Giedt},\ and\ \citenamefont
  {Greensite}}]{Ambjorn:1999ym}%
  \BibitemOpen
  \bibfield  {author} {\bibinfo {author} {\bibfnamefont {J.}~\bibnamefont
  {Ambjorn}}, \bibinfo {author} {\bibfnamefont {J.}~\bibnamefont {Giedt}},\
  and\ \bibinfo {author} {\bibfnamefont {J.}~\bibnamefont {Greensite}},\
  }\bibfield  {title} {\bibinfo {title} {{Vortex structure vs. monopole
  dominance in Abelian-projected gauge theory}},\ }\href
  {https://doi.org/10.1088/1126-6708/2000/02/033} {\bibfield  {journal}
  {\bibinfo  {journal} {J. High Energy Phys.}\ }\textbf {\bibinfo {volume}
  {02}},\ \bibinfo {pages} {033}},\ \Eprint
  {https://arxiv.org/abs/hep-lat/9907021} {arXiv:hep-lat/9907021} \BibitemShut
  {NoStop}%
\bibitem [{\citenamefont {Alexandrou}\ \emph {et~al.}(2000)\citenamefont
  {Alexandrou}, \citenamefont {de~Forcrand},\ and\ \citenamefont
  {D'Elia}}]{Alexandrou:1999vx}%
  \BibitemOpen
  \bibfield  {author} {\bibinfo {author} {\bibfnamefont {C.}~\bibnamefont
  {Alexandrou}}, \bibinfo {author} {\bibfnamefont {P.}~\bibnamefont
  {de~Forcrand}},\ and\ \bibinfo {author} {\bibfnamefont {M.}~\bibnamefont
  {D'Elia}},\ }\bibfield  {title} {\bibinfo {title} {{The role of center
  vortices in QCD}},\ }\href {https://doi.org/10.1016/S0375-9474(99)00763-0}
  {\bibfield  {journal} {\bibinfo  {journal} {Nucl. Phys. A}\ }\textbf
  {\bibinfo {volume} {663}},\ \bibinfo {pages} {1031} (\bibinfo {year}
  {2000})},\ \Eprint {https://arxiv.org/abs/hep-lat/9909005}
  {arXiv:hep-lat/9909005} \BibitemShut {NoStop}%
\bibitem [{\citenamefont {de~Forcrand}\ and\ \citenamefont
  {Pepe}(2001)}]{deForcrand:2000pg}%
  \BibitemOpen
  \bibfield  {author} {\bibinfo {author} {\bibfnamefont {P.}~\bibnamefont
  {de~Forcrand}}\ and\ \bibinfo {author} {\bibfnamefont {M.}~\bibnamefont
  {Pepe}},\ }\bibfield  {title} {\bibinfo {title} {{Center vortices and
  monopoles without lattice Gribov copies}},\ }\href
  {https://doi.org/10.1016/S0550-3213(01)00009-8} {\bibfield  {journal}
  {\bibinfo  {journal} {Nucl. Phys. B}\ }\textbf {\bibinfo {volume} {598}},\
  \bibinfo {pages} {557} (\bibinfo {year} {2001})},\ \Eprint
  {https://arxiv.org/abs/hep-lat/0008016} {arXiv:hep-lat/0008016} \BibitemShut
  {NoStop}%
\bibitem [{\citenamefont {Reinhardt}(2002)}]{Reinhardt:2001kf}%
  \BibitemOpen
  \bibfield  {author} {\bibinfo {author} {\bibfnamefont {H.}~\bibnamefont
  {Reinhardt}},\ }\bibfield  {title} {\bibinfo {title} {{Topology of center
  vortices}},\ }\href {https://doi.org/10.1016/S0550-3213(02)00130-X}
  {\bibfield  {journal} {\bibinfo  {journal} {Nucl. Phys. B}\ }\textbf
  {\bibinfo {volume} {628}},\ \bibinfo {pages} {133} (\bibinfo {year}
  {2002})},\ \Eprint {https://arxiv.org/abs/hep-th/0112215}
  {arXiv:hep-th/0112215} \BibitemShut {NoStop}%
\bibitem [{\citenamefont {Oxman}(2018)}]{Oxman:2018dzp}%
  \BibitemOpen
  \bibfield  {author} {\bibinfo {author} {\bibfnamefont {L.~E.}\ \bibnamefont
  {Oxman}},\ }\bibfield  {title} {\bibinfo {title} {{4D ensembles of
  percolating center vortices and monopole defects: The emergence of flux tubes
  with $N$-ality and gluon confinement}},\ }\href
  {https://doi.org/10.1103/PhysRevD.98.036018} {\bibfield  {journal} {\bibinfo
  {journal} {Phys. Rev. D}\ }\textbf {\bibinfo {volume} {98}},\ \bibinfo
  {pages} {036018} (\bibinfo {year} {2018})},\ \Eprint
  {https://arxiv.org/abs/1805.06354} {arXiv:1805.06354 [hep-th]} \BibitemShut
  {NoStop}%
\bibitem [{\citenamefont {Junior}\ \emph {et~al.}(2022)\citenamefont {Junior},
  \citenamefont {Oxman},\ and\ \citenamefont {Reinhardt}}]{Junior:2022bol}%
  \BibitemOpen
  \bibfield  {author} {\bibinfo {author} {\bibfnamefont {D.~R.}\ \bibnamefont
  {Junior}}, \bibinfo {author} {\bibfnamefont {L.~E.}\ \bibnamefont {Oxman}},\
  and\ \bibinfo {author} {\bibfnamefont {H.}~\bibnamefont {Reinhardt}},\
  }\bibfield  {title} {\bibinfo {title} {{Infrared Yang-Mills wave functional
  due to percolating center vortices}},\ }\href
  {https://doi.org/10.1103/PhysRevD.106.114021} {\bibfield  {journal} {\bibinfo
   {journal} {Phys. Rev. D}\ }\textbf {\bibinfo {volume} {106}},\ \bibinfo
  {pages} {114021} (\bibinfo {year} {2022})},\ \Eprint
  {https://arxiv.org/abs/2211.03006} {arXiv:2211.03006 [hep-th]} \BibitemShut
  {NoStop}%
\bibitem [{\citenamefont {Junior}\ and\ \citenamefont
  {Oxman}(2025)}]{Junior:2024urr}%
  \BibitemOpen
  \bibfield  {author} {\bibinfo {author} {\bibfnamefont {D.~R.}\ \bibnamefont
  {Junior}}\ and\ \bibinfo {author} {\bibfnamefont {L.~E.}\ \bibnamefont
  {Oxman}},\ }\bibfield  {title} {\bibinfo {title} {{Ensembles of center
  vortices and chains: Insights from a natural lattice framework}},\ }\href
  {https://doi.org/10.1103/PhysRevD.111.054036} {\bibfield  {journal} {\bibinfo
   {journal} {Phys. Rev. D}\ }\textbf {\bibinfo {volume} {111}},\ \bibinfo
  {pages} {054036} (\bibinfo {year} {2025})},\ \Eprint
  {https://arxiv.org/abs/2411.04325} {arXiv:2411.04325 [hep-th]} \BibitemShut
  {NoStop}%
\bibitem [{\citenamefont {Montero}(1999)}]{Montero:1999by}%
  \BibitemOpen
  \bibfield  {author} {\bibinfo {author} {\bibfnamefont {A.}~\bibnamefont
  {Montero}},\ }\bibfield  {title} {\bibinfo {title} {{Study of SU(3)
  vortex-like configurations with a new maximal center gauge fixing method}},\
  }\href {https://doi.org/10.1016/S0370-2693(99)01113-2} {\bibfield  {journal}
  {\bibinfo  {journal} {Phys. Lett. B}\ }\textbf {\bibinfo {volume} {467}},\
  \bibinfo {pages} {106} (\bibinfo {year} {1999})},\ \Eprint
  {https://arxiv.org/abs/hep-lat/9906010} {arXiv:hep-lat/9906010} \BibitemShut
  {NoStop}%
\bibitem [{\citenamefont {Faber}\ \emph
  {et~al.}(1999{\natexlab{b}})\citenamefont {Faber}, \citenamefont {Greensite},
  \citenamefont {Olejn{\'i}k},\ and\ \citenamefont {Yamada}}]{Faber:1999gu}%
  \BibitemOpen
  \bibfield  {author} {\bibinfo {author} {\bibfnamefont {M.}~\bibnamefont
  {Faber}}, \bibinfo {author} {\bibfnamefont {J.}~\bibnamefont {Greensite}},
  \bibinfo {author} {\bibfnamefont {{\v S}.}~\bibnamefont {Olejn{\'i}k}},\ and\
  \bibinfo {author} {\bibfnamefont {D.}~\bibnamefont {Yamada}},\ }\bibfield
  {title} {\bibinfo {title} {{The vortex-finding property of maximal center
  (and other) gauges}},\ }\href {https://doi.org/10.1088/1126-6708/1999/12/012}
  {\bibfield  {journal} {\bibinfo  {journal} {J. High Energy Phys.}\ }\textbf
  {\bibinfo {volume} {12}},\ \bibinfo {pages} {012}},\ \Eprint
  {https://arxiv.org/abs/hep-lat/9910033} {arXiv:hep-lat/9910033} \BibitemShut
  {NoStop}%
\bibitem [{\citenamefont {Creutz}(1980)}]{Creutz:1980zw}%
  \BibitemOpen
  \bibfield  {author} {\bibinfo {author} {\bibfnamefont {M.}~\bibnamefont
  {Creutz}},\ }\bibfield  {title} {\bibinfo {title} {{Monte Carlo study of
  quantized SU(2) gauge theory}},\ }\href
  {https://doi.org/10.1103/PhysRevD.21.2308} {\bibfield  {journal} {\bibinfo
  {journal} {Phys. Rev. D}\ }\textbf {\bibinfo {volume} {21}},\ \bibinfo
  {pages} {2308} (\bibinfo {year} {1980})}\BibitemShut {NoStop}%
\bibitem [{\citenamefont {Wilson}(1974)}]{Wilson:1974sk}%
  \BibitemOpen
  \bibfield  {author} {\bibinfo {author} {\bibfnamefont {K.~G.}\ \bibnamefont
  {Wilson}},\ }\bibfield  {title} {\bibinfo {title} {{Confinement of quarks}},\
  }\href {https://doi.org/10.1103/PhysRevD.10.2445} {\bibfield  {journal}
  {\bibinfo  {journal} {Phys. Rev. D}\ }\textbf {\bibinfo {volume} {10}},\
  \bibinfo {pages} {2445} (\bibinfo {year} {1974})}\BibitemShut {NoStop}%
\bibitem [{\citenamefont {Cabibbo}\ and\ \citenamefont
  {Marinari}(1982)}]{Cabibbo:1982zn}%
  \BibitemOpen
  \bibfield  {author} {\bibinfo {author} {\bibfnamefont {N.}~\bibnamefont
  {Cabibbo}}\ and\ \bibinfo {author} {\bibfnamefont {E.}~\bibnamefont
  {Marinari}},\ }\bibfield  {title} {\bibinfo {title} {{A new method for
  updating SU($N$) matrices in computer simulations of gauge theories}},\
  }\href {https://doi.org/10.1016/0370-2693(82)90696-7} {\bibfield  {journal}
  {\bibinfo  {journal} {Phys. Lett. B}\ }\textbf {\bibinfo {volume} {119}},\
  \bibinfo {pages} {387} (\bibinfo {year} {1982})}\BibitemShut {NoStop}%
\bibitem [{\citenamefont {Lucini}\ \emph {et~al.}(2004)\citenamefont {Lucini},
  \citenamefont {Teper},\ and\ \citenamefont {Wenger}}]{Lucini:2004my}%
  \BibitemOpen
  \bibfield  {author} {\bibinfo {author} {\bibfnamefont {B.}~\bibnamefont
  {Lucini}}, \bibinfo {author} {\bibfnamefont {M.}~\bibnamefont {Teper}},\ and\
  \bibinfo {author} {\bibfnamefont {U.}~\bibnamefont {Wenger}},\ }\bibfield
  {title} {\bibinfo {title} {{Glueballs and k-strings in SU($N$) gauge
  theories: calculations with improved operators}},\ }\href
  {https://doi.org/10.1088/1126-6708/2004/06/012} {\bibfield  {journal}
  {\bibinfo  {journal} {J. High Energy Phys.}\ }\textbf {\bibinfo {volume}
  {06}},\ \bibinfo {pages} {012}},\ \Eprint
  {https://arxiv.org/abs/hep-lat/0404008} {arXiv:hep-lat/0404008} \BibitemShut
  {NoStop}%
\bibitem [{\citenamefont {Lucini}\ \emph {et~al.}(2005)\citenamefont {Lucini},
  \citenamefont {Teper},\ and\ \citenamefont {Wenger}}]{Lucini:2005vg}%
  \BibitemOpen
  \bibfield  {author} {\bibinfo {author} {\bibfnamefont {B.}~\bibnamefont
  {Lucini}}, \bibinfo {author} {\bibfnamefont {M.}~\bibnamefont {Teper}},\ and\
  \bibinfo {author} {\bibfnamefont {U.}~\bibnamefont {Wenger}},\ }\bibfield
  {title} {\bibinfo {title} {{Properties of the deconfining phase transition in
  SU($N$) gauge theories}},\ }\href
  {https://doi.org/10.1088/1126-6708/2005/02/033} {\bibfield  {journal}
  {\bibinfo  {journal} {J. High Energy Phys.}\ }\textbf {\bibinfo {volume}
  {02}},\ \bibinfo {pages} {033}},\ \Eprint
  {https://arxiv.org/abs/hep-lat/0502003} {arXiv:hep-lat/0502003} \BibitemShut
  {NoStop}%
\bibitem [{\citenamefont {Bali}\ \emph {et~al.}(1997)\citenamefont {Bali},
  \citenamefont {Schilling},\ and\ \citenamefont {Wachter}}]{Bali:1997am}%
  \BibitemOpen
  \bibfield  {author} {\bibinfo {author} {\bibfnamefont {G.~S.}\ \bibnamefont
  {Bali}}, \bibinfo {author} {\bibfnamefont {K.}~\bibnamefont {Schilling}},\
  and\ \bibinfo {author} {\bibfnamefont {A.}~\bibnamefont {Wachter}},\
  }\bibfield  {title} {\bibinfo {title} {{Complete $O(v^2)$ corrections to the
  static interquark potential from SU(3) gauge theory}},\ }\href
  {https://doi.org/10.1103/PhysRevD.56.2566} {\bibfield  {journal} {\bibinfo
  {journal} {Phys. Rev. D}\ }\textbf {\bibinfo {volume} {56}},\ \bibinfo
  {pages} {2566} (\bibinfo {year} {1997})},\ \Eprint
  {https://arxiv.org/abs/hep-lat/9703019} {arXiv:hep-lat/9703019} \BibitemShut
  {NoStop}%
\bibitem [{\citenamefont {Edwards}\ \emph {et~al.}(1998)\citenamefont
  {Edwards}, \citenamefont {Heller},\ and\ \citenamefont
  {Klassen}}]{Edwards:1997xf}%
  \BibitemOpen
  \bibfield  {author} {\bibinfo {author} {\bibfnamefont {R.~G.}\ \bibnamefont
  {Edwards}}, \bibinfo {author} {\bibfnamefont {U.~M.}\ \bibnamefont
  {Heller}},\ and\ \bibinfo {author} {\bibfnamefont {T.~R.}\ \bibnamefont
  {Klassen}},\ }\bibfield  {title} {\bibinfo {title} {{Accurate scale
  determinations for the Wilson gauge action}},\ }\href
  {https://doi.org/10.1016/S0550-3213(98)80003-5} {\bibfield  {journal}
  {\bibinfo  {journal} {Nucl. Phys. B}\ }\textbf {\bibinfo {volume} {517}},\
  \bibinfo {pages} {377} (\bibinfo {year} {1998})},\ \Eprint
  {https://arxiv.org/abs/hep-lat/9711003} {arXiv:hep-lat/9711003} \BibitemShut
  {NoStop}%
\bibitem [{\citenamefont {Allton}\ \emph {et~al.}(2008)\citenamefont {Allton},
  \citenamefont {Teper},\ and\ \citenamefont {Trivini}}]{Allton:2008ty}%
  \BibitemOpen
  \bibfield  {author} {\bibinfo {author} {\bibfnamefont {C.}~\bibnamefont
  {Allton}}, \bibinfo {author} {\bibfnamefont {M.}~\bibnamefont {Teper}},\ and\
  \bibinfo {author} {\bibfnamefont {A.}~\bibnamefont {Trivini}},\ }\bibfield
  {title} {\bibinfo {title} {{On the running of the bare coupling in SU($N$)
  lattice gauge theories}},\ }\href
  {https://doi.org/10.1088/1126-6708/2008/07/021} {\bibfield  {journal}
  {\bibinfo  {journal} {J. High Energy Phys.}\ }\textbf {\bibinfo {volume}
  {07}},\ \bibinfo {pages} {021}},\ \Eprint {https://arxiv.org/abs/0803.1092}
  {arXiv:0803.1092 [hep-lat]} \BibitemShut {NoStop}%
\bibitem [{\citenamefont {Athenodorou}\ and\ \citenamefont
  {Teper}(2020)}]{Athenodorou:2020ani}%
  \BibitemOpen
  \bibfield  {author} {\bibinfo {author} {\bibfnamefont {A.}~\bibnamefont
  {Athenodorou}}\ and\ \bibinfo {author} {\bibfnamefont {M.}~\bibnamefont
  {Teper}},\ }\bibfield  {title} {\bibinfo {title} {{The glueball spectrum of
  SU(3) gauge theory in 3 + 1 dimensions}},\ }\href
  {https://doi.org/10.1007/JHEP11(2020)172} {\bibfield  {journal} {\bibinfo
  {journal} {J. High Energy Phys.}\ }\textbf {\bibinfo {volume} {11}},\
  \bibinfo {pages} {172}},\ \Eprint {https://arxiv.org/abs/2007.06422}
  {arXiv:2007.06422 [hep-lat]} \BibitemShut {NoStop}%
\bibitem [{\citenamefont {Athenodorou}\ and\ \citenamefont
  {Teper}(2021)}]{Athenodorou:2021qvs}%
  \BibitemOpen
  \bibfield  {author} {\bibinfo {author} {\bibfnamefont {A.}~\bibnamefont
  {Athenodorou}}\ and\ \bibinfo {author} {\bibfnamefont {M.}~\bibnamefont
  {Teper}},\ }\bibfield  {title} {\bibinfo {title} {{SU(N) gauge theories in 3
  + 1 dimensions: glueball spectrum, string tensions and topology}},\ }\href
  {https://doi.org/10.1007/JHEP12(2021)082} {\bibfield  {journal} {\bibinfo
  {journal} {J. High Energy Phys.}\ }\textbf {\bibinfo {volume} {12}},\
  \bibinfo {pages} {082}},\ \Eprint {https://arxiv.org/abs/2106.00364}
  {arXiv:2106.00364 [hep-lat]} \BibitemShut {NoStop}%
\end{thebibliography}%

\onecolumngrid\clearpage\twocolumngrid

\title{Supplemental material:\ Structure of center vortex matter in SU(4) Yang-Mills theory}
\begin{abstract}
This supplementary document provides interactive 3D models and animations of the center vortex structure over three-dimensional slices of the lattice, expanding on the static images in the main text. The animations are produced using the \texttt{animate} package in \LaTeX. To interact with the models and animations, readers must open this document in Adobe Acrobat Reader.

``Multimedia \& 3D" must be enabled in Acrobat Reader for the interactive models to be available, and enabling ``double-sided rendering" is also necessary for proper rendering. To activate the models, simply click on the image. To rotate the model, click and hold the left mouse button and move the mouse. Use the scroll wheel or shift-click to zoom. Preset views of the model focused on faces of the three-dimensional volume can be accessed by right clicking and using the ``Views" menu.

The controls for the animations are located below their thumbnails. From left to right, these are: stop and go to first frame, step backwards one frame, play backwards, play forwards, step forwards one frame, and stop and go to last frame. Simply click on the desired control to interact with the animation. After clicking either play button, they will be replaced with a pause button which can subsequently be used to stop the animation.
\end{abstract}

\makeatletter
\@booleanfalse\preprint@sw
\makeatother
\maketitle

\setcounter{figure}{0}
\renewcommand\thefigure{S-\arabic{figure}}
\renewcommand{\theHfigure}{Supplement.\thefigure}

\begin{figure*}
	\centering
	\includemedia[noplaybutton,3Dtoolbar,3Dmenu,label=coarse,3Dviews=coarse_views.vws, width=0.88\textwidth]{\includegraphics{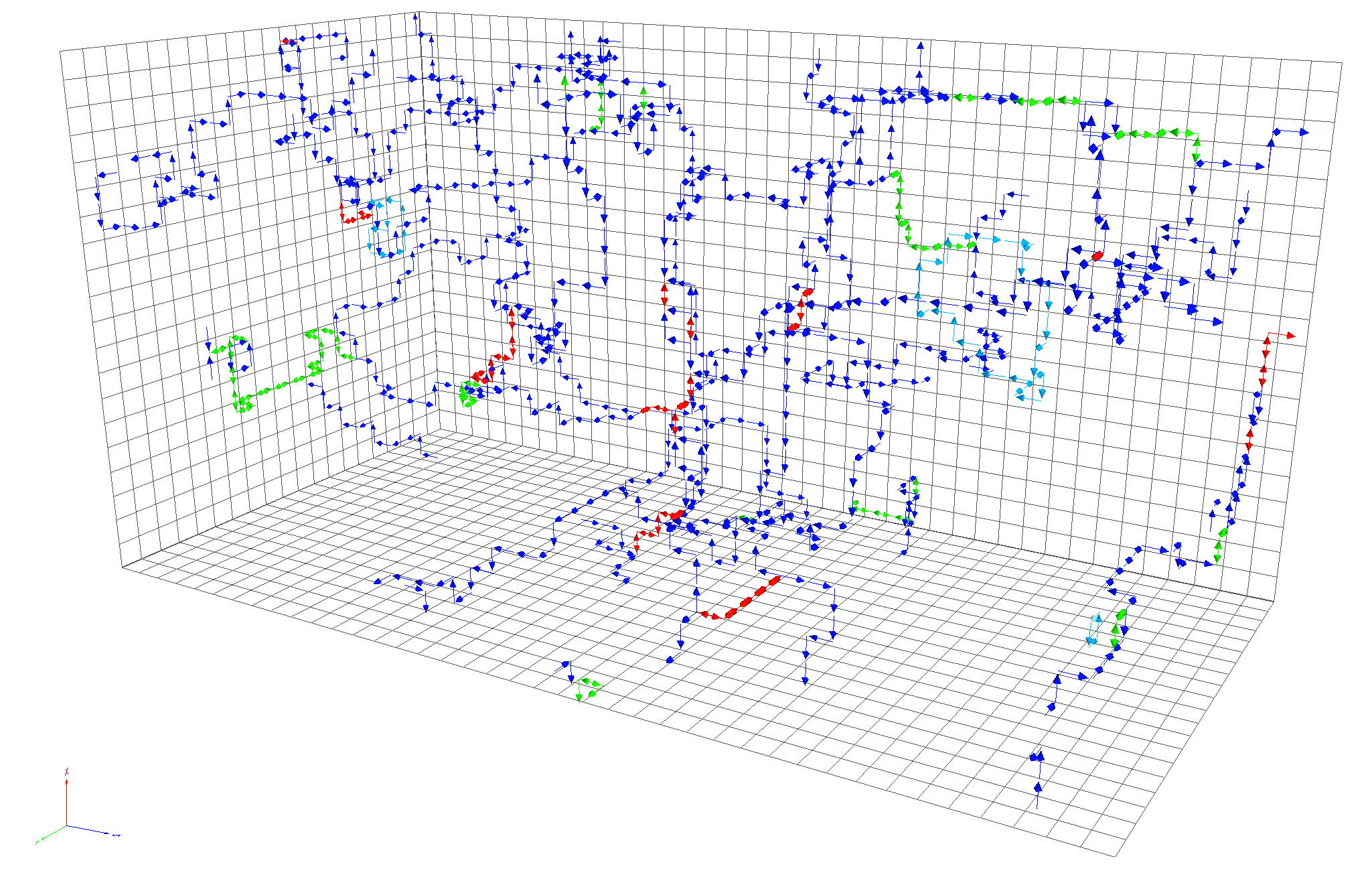}}{coarse.u3d}
	
	\vspace{2.5em}
	
	\animategraphics[loop,nomouse,width=0.88\linewidth,controls={play,stop,step}]{4}{coarse_animation/y}{1}{20}
	\caption{\label{fig:coarse_supp} Interactive graphic (\textbf{top}) and animation (\textbf{bottom}) of the $\mathrm{SU}(4)$ center vortex structure in spatial slices of the lattice at $a\sqrt{\sigma} \simeq 0.239$. The three-dimensional volume is $20 \times 20 \times 40$.}
\end{figure*}

\begin{figure*}
	\centering
	\includemedia[noplaybutton,3Dtoolbar,3Dmenu,label=medium,3Dviews=medium_views.vws, width=0.88\textwidth]{\includegraphics{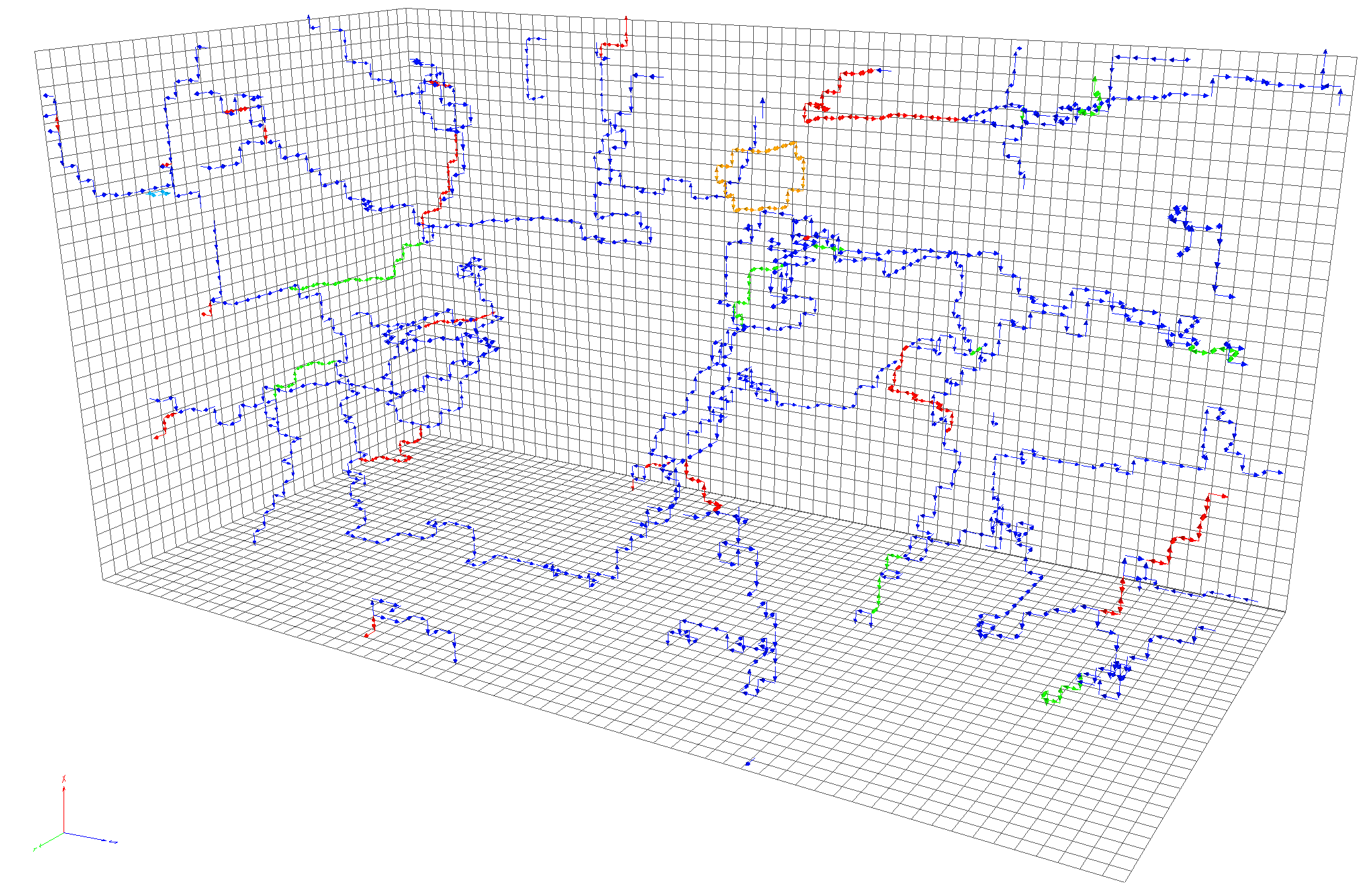}}{medium.u3d}
	
	\vspace{2.5em}
	
	\animategraphics[loop,nomouse,width=0.88\linewidth,controls={play,stop,step}]{4}{medium_animation/z}{1}{32}
	\caption{\label{fig:medium_supp} Interactive graphic (\textbf{top}) and animation (\textbf{bottom}) of the $\mathrm{SU}(4)$ center vortex structure in spatial slices of the lattice at $a\sqrt{\sigma} \simeq 0.149$. The three-dimensional volume is $32 \times 32 \times 64$.}
\end{figure*}

\begin{figure*}
	\centering
	\includemedia[noplaybutton,3Dtoolbar,3Dmenu,label=fine,3Dviews=fine_views.vws, width=0.88\textwidth]{\includegraphics{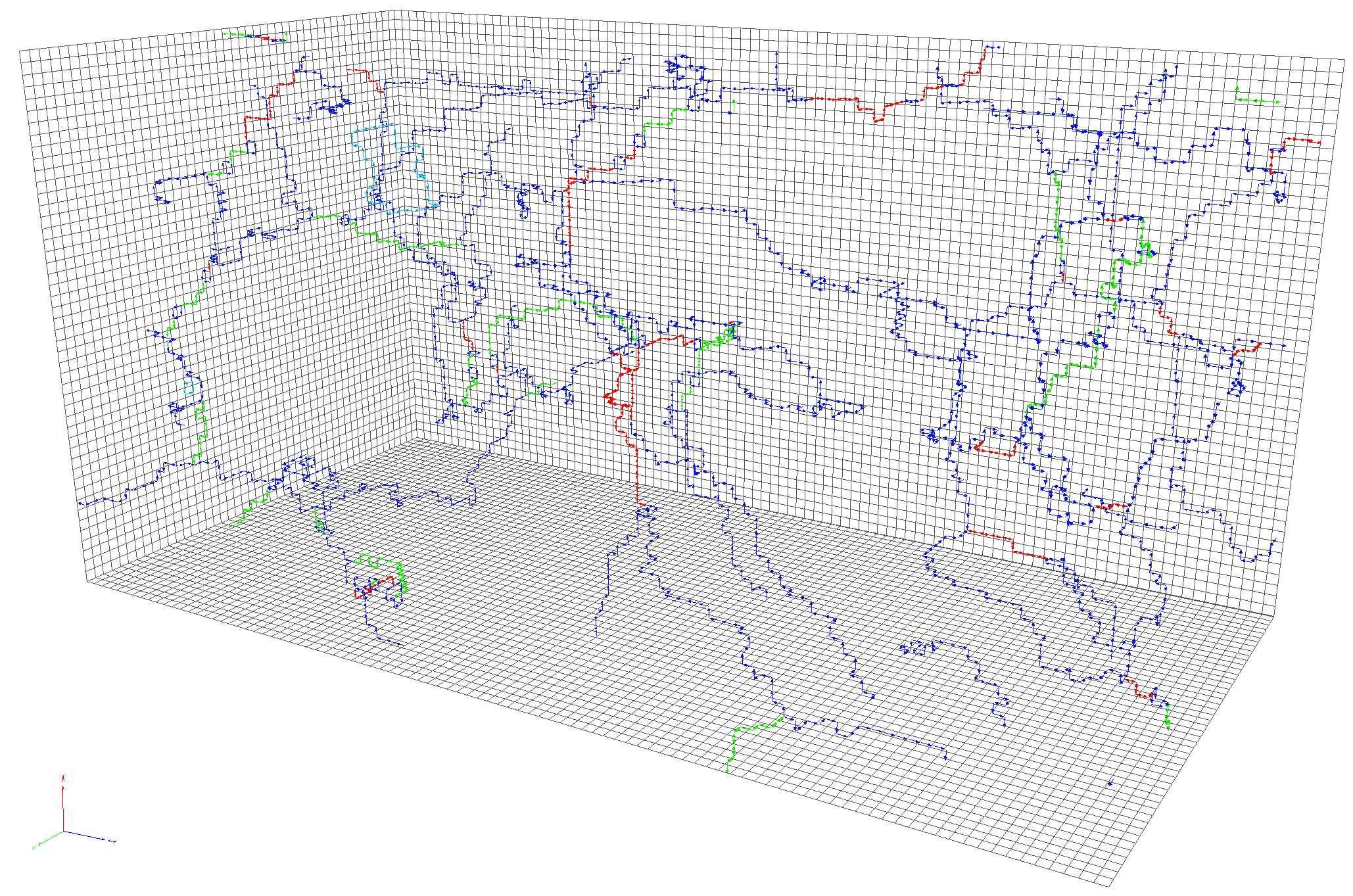}}{fine.u3d}
	
	\vspace{2.5em}
	
	\animategraphics[loop,nomouse,width=0.88\linewidth,controls={play,stop,step}]{4}{fine_animation/z}{1}{48}
	\caption{\label{fig:fine_supp} Interactive graphic (\textbf{top}) and animation (\textbf{bottom}) of the $\mathrm{SU}(4)$ center vortex structure in spatial slices of the lattice at $a\sqrt{\sigma} \simeq 0.103$. The three-dimensional volume is $48 \times 48 \times 96$.}
\end{figure*}

\end{document}